\documentclass[useAMS,usenatbib]{mn2e}

\usepackage{graphicx}
\usepackage{amsmath,amssymb}

\def\apjl{ApJL}
\def\apjs{ApJS}
\def\aap{A\&A}

\title[Equilibrium Model]{Equilibrium Model Constraints on Baryon Cycling Across Cosmic Time}

\author[Mitra, Dav\'e \& Finlator]{
\parbox[t]{\textwidth}{\vspace{-1cm}
Sourav Mitra$^{1}$\thanks{E-mail: hisourav@gmail.com},
Romeel Dav\'e$^{1,2,3,4}$,
Kristian Finlator$^{5}$}
\\
\\$^1$ University of the Western Cape, Bellville, Cape Town 7535, South Africa
\\$^2$ South African Astronomical Observatories, Observatory, Cape Town 7925, South Africa
\\$^3$ African Institute for Mathematical Sciences, Muizenberg, Cape Town 7945, South Africa
\\$^4$ Astronomy Department, University of Arizona, Tucson, AZ 85721, USA
\\$^5$ DARK fellow, Dark Cosmology Centre, Niels Bohr Institute, University of Copenhagen
}

\begin{document}

\maketitle

 \begin{abstract}
Galaxies strongly self-regulate their growth via energetic feedback
from stars, supernovae, and black holes, but these processes are
among the least understood aspects of galaxy formation theory.  We
present an analytic galaxy evolution model that directly constrains
such feedback processes from observed galaxy scaling relations. The
equilibrium model, which is broadly valid for star-forming central galaxies
that dominate cosmic star formation,
is based on the ansatz that galaxies live in a slowly-evolving equilibrium
between inflows, outflows, and star formation. Using a Bayesian
Monte Carlo Markov chain approach, we constrain our model to match
observed galaxy scaling relations between stellar mass and halo
mass, star formation rate, and metallicity from $0<z<2$.  A good
fit ($\chi^2\approx 1.6$) is achieved with eight free parameters.
We further show that constraining our model to any two of the three
data sets also produces a fit to the third that is within reasonable
systematic uncertainties.  The resulting best-fit parameters that
describe baryon cycling suggest galactic outflow scalings intermediate
between energy and momentum-driven winds, a weak dependence of wind
recycling time on mass, and a quenching mass scale that evolves
modestly upwards with redshift. This model further predicts a stellar
mass-star formation rate relation that is in good agreement with
observations to $z\sim 6$.  Our results suggest that this simple
analytic framework captures the basic physical processes required
to model the mean evolution of stars and metals in galaxies, despite
not incorporating many canonical ingredients of galaxy formation
models such as merging or disk formation.
\end{abstract}


\section{Introduction}

The formation and evolution of galaxies is governed by a vast number
of interrelated physical processes that span many decades in physical
and temporal scale.  Such complexity makes this area of astrophysics
extraordinarily rich and interesting, driven by the accelerating
pace of observational data that has statistically characterised the
multi-wavelength properties of galaxies from locally out to high
redshifts.  Yet, this physical complexity and wealth of data has
proved to be an increasingly difficult challenge for models of
galaxy formation, whose primary goal is to sort out which processes
are responsible for which observables, and thereby elucidate the
physics that drives galaxy evolution.

Currently, there are two main approaches to modeling galaxy formation:
semi-analytic models (SAMs) and hydrodynamic simulations.  SAMs are
rooted in the classical analytic approaches of the 1970's
\citep{1977MNRAS.179..541R,1977ApJ...215..483B,1978MNRAS.183..341W}, which
posit that galaxies form as central baryonic condensations cooled
from the virial temperature within dark matter halos, conserving
angular momentum to form a disk
\citep{1980MNRAS.193..189F,1997ApJ...482..659D,1998MNRAS.295..319M}.  With
the emergence of the hierarchical cold dark matter (CDM) paradigm
\citep{1984Natur.311..517B}, such disks followed the merging of
dark matter halos via ``merger trees" to build the observed galaxy
population
\citep{1991ApJ...379...52W,1993MNRAS.264..201K,1999MNRAS.310.1087S}.  As
the galaxy population became characterised with greater accuracy
over a wider range of epochs, numerous free parameters were layered
upon this framework in order to reproduce various observed galaxy
properties, with these free parameters constrained by additional,
putatively independent, observations.

Today, SAMs typically employ $\sim 30-50$ parameters, the majority
of which characterise the gas dynamics within and around halos,
since that is not tracked directly in these models.  With such a
large number of parameters, the robustness of the physical intuition
gained from such models becomes difficult to ascertain.
Not only the number of parameters, but also the fact that many more
equations are needed in order to capture various physical phenomena
makes it enormously complex.
Numerous different SAMs all match a similarly broad set of data, but often
do so by invoking substantively different physical recipes and
parameterizations \citep{Somerville:2015}. To mitigate this, SAMs
have begun employing more sophisticated statistical approaches such
as Monte Carlo Markov Chain (MCMC) parameter space searches that
use Bayesian inference to obtain probability distributions and
correlations.  However, it is computationally infeasible to apply
this approach to the entire parameter set, and thus it is typically
only applied to a subset of parameters or a set of available observations
\citep{2012MNRAS.421.1779L,2013MNRAS.431.3373H} or a ``stripped-down"
SAM \citep{2014MNRAS.444.2599B}.

Hydrodynamic simulations have been gaining rapidly in popularity
and importance, owing to advances in computing and input physics.
Such calculations remove the need for the vast majority of SAM
parameter that describe the gas dynamics, but typically still require
parameters to describe ``sub-grid" processes such as star formation
and feedback.  These days, simulations can reproduce a similarly
wide range of data as SAMs
\citep[e.g.][]{dav11a,2014Natur.509..177V,2015MNRAS.446..521S}.
While impressive, the operational complexity and computational 
cost of such models makes it
less than straightforward to interpret the simulations in order to
robustly connect physical processes with observables.  Furthermore,
a state-of-the-art simulation typically require millions of CPU-hours
on a supercomputer, making extensive parameter space explorations
prohibitive.

SAMs and simulations complement and inform each other, but the
complexity of each approach present different hurdles for reaching
the ultimate goal, namely a simple and intuitive understanding of
the underlying physics driving galaxy evolution.  As a result, the
prevailing intuition in the galaxy formation community remains
rooted in decades-old analytic models, even as simulations argue
for significant revisions such as that star formation is fueled
primarily by cold, smooth accretion \citep[e.g.][]{2005MNRAS.363....2K}
and that galactic outflows drive the angular momentum distribution
within disks \citep[e.g.][]{2012MNRAS.419..771B}, and even as
observations question key aspects of the most basic paradigm such
as the existence of ubiquitous large hot halos of gas around
star-forming spiral galaxies \citep{2001MNRAS.320..261B,2009MNRAS.399.1773C}.
Hence despite great progress, it is becoming increasingly difficult
to separate which physical processes in galaxies (and their surrounding
gas) are crucial for establishing which observable properties and
their evolution.

The advent of large multi-wavelength observational surveys has
afforded us new opportunities to study the physical properties of
galaxies in greater statistical detail.  Importantly, such surveys
have enabled the robust characterisation of {\it scaling relations}
between physical properties of galaxies such as their stars, gas,
dust, metals, black holes, and dark matter.  Such scaling relations
are often more directly relatable to underlying physical processes
than simple counting statistics.  One example is the relationship
between star formation rate and stellar mass ($M_*$), known as the
``main sequence" \citep{2007ApJ...660L..43N}; the tightness of this
relation observed out to high redshifts implies that galaxies grow
mostly steadily rather than via bursts \citep[e.g.][]{2015A&A...575A..74S}.
Another example is the relationship between $M_*$ and gas-phase
metallicity ($Z$), which with its scatter of $\sim 0.1$ dex is among
the tightest galaxy scaling relation known~\citep{2004ApJ...613..898T}.
While current SAMs and simulations broadly succeed at reproducing these relations and their
evolution~\citep{2014arXiv1410.0365H,2015MNRAS.446..521S,Somerville:2015}, the physical
interpretation through the lens of highly complex models with many
free parameters remains less than straightforward.

A recent addition to the cadre of physical models of galaxy formation
is what we refer to as an ``equilibrium model"
\citep{2008MNRAS.385.2181F,dav12}.  This is similar to a ``bathtub
model" \citep{2010ApJ...718.1001B,2014MNRAS.444.2071D} or a ``gas
regulator model" \citep{2013ApJ...772..119L,2014MNRAS.443.3643P},
though there are non-trivial differences among these.  While also
analytically-based, an equilibrium model is quite different than a
SAM; it does not employ a merger tree or the idea of cooling gas
onto a disk via angular momentum conservation.  Instead, an equilibrium
model begins with a simple mass balance equation in the interstellar
medium (ISM) that the inflows and outflows balance on a timescale which
is very short compared to a Hubble time, with a non-trivial implication
that many of the time evolution terms governing galaxy formation can be set to zero.
Inflow into the ISM owes, as suggested by hydrodynamic
simulations, predominantly to gravitationally-driven smooth accretion.
Continual outflows eject gas from the ISM, and are incorporated as
a core aspect rather than as an epicyclic add-on, as is preventive
feedback owing to photoionisation or active galactic nuclei.
Similarly, recycling of some outflow material back into the ISM is
also a core aspect
\citep{2010MNRAS.406.2325O,2012MNRAS.422.2816B,2013MNRAS.431.3373H}. The
parameters of the equilibrium model thus describe the motion of gas
into and out of galaxies, which has been dubbed the {\it baryon cycle.}

The equilibrium model in particular makes the additional assumption
that the gas reservoir in a galaxy is slowly evolving.  This is
motivated by hydrodynamic simulations \citep{2008MNRAS.385.2181F}
as well as simple theoretical expectations
\citep{2010ApJ...718.1001B,2014MNRAS.444.2071D}, and is in accord
with observations of gas content over cosmological timescales
\citep{2013ApJ...778....2S}.  However, this is clearly not true on
shorter timescales, and for instance the gas regulator model explicitly
relaxes this assumption \citep{2013ApJ...772..119L}. The gas regulator
model is thus more general, but its equations are significantly more
complex, with a commensurate reduction in the ease of interpretation.
Nonetheless it is also an instructive approach, but here for clarity
we will not explicitly consider the gas reservoir, and leave it for
future work.

While simple and intuitive, the baryon cycle-driven scenario forwarded
by the equilibrium model and its cousins is at face value substantively
different than the merger-driven, halo-centric view of galaxies
that underlies the canonical paradigm.  This prevailing view has
been quite successful at reproducing observations, albeit requiring
dozens of free parameters to do so.  A natural question then arises,
can an equilibrium model also reproduce basic properties the galaxy
population as observed across cosmic time? The equilibrium model
has been explicitly implemented to interpret the measurements of
mass-metallicity (MZ) relation \citep{2013ApJ...769..148H},
where they found a relative tension between
the predicted MZ relation and star formation rates (SFRs) of low mass galaxies
for some simple choice of baryon cycling parameters. This also motivates
to ask whether the equilibrium model can directly accommodate those observations
by considering these parameters as free.

In this paper, we examine how well a simple equilibrium model can
match observations of key galaxy scaling relations from today until
``cosmic noon" at redshift $z\sim 2$.  We represent the baryon
cycling parameters with free variables that we constrain against
observations of the halo mass-stellar mass and mass-metallicity
relations from $z=0-2$, using a Bayesian MCMC approach.  We show
that the equilibrium model is able to match these observations
within acceptable statistical bounds, using eight free parameters (minimum
number we justify using the Bayesian evidence). Moreover, the same
model is then able to predict the evolution of the SFR-$M_*$ (the
main sequence) from $z=0-2$ in good agreement with observations,
which at $z\sim 2$ has been difficult to achieve in other models.
The resulting constraints on the baryon cycling parameters provide
new insights into the physical processes that govern ejective and
preventive feedback in galaxies over the majority of cosmic time.
More broadly, our results suggest that the baryon cycling-based,
mass-balance paradigm is a viable representation of the mean growth
of stars and metals in galaxies over cosmological timescales, and
hence the equilibrium model provides a novel and interesting framework
for constructing more detailed models of galaxy evolution that
may provide clearer intuition about its physics drivers.

This paper is organised as follows. In Section \ref{sec:model}, we
introduce the equilibrium model, and our chosen set of parameterisations
for the baryon cycling parameters. In Section \ref{sec:fits}
and  \ref{sec:Bayesian}, we
present the best fits to the observed scaling relations obtained
via our MCMC, and present Bayesian evidence arguments for why these
particular parameters are necessary and sufficient. In Section
\ref{sec:implications}, we discuss our model predictions for higher
redshifts and the general evolution of stellar and metal content
of galaxies at various masses. Finally, we summarise and discuss
our results more broadly in Section \ref{sec:summary}.
Throughout this paper we assume a flat $\Lambda$CDM cosmology with
$\Omega_m=1-\Omega_\Lambda=0.27$ and $h=0.7$ \citep{2011ApJS..192...16L}.

\section{Methods}\label{sec:model}

\subsection{The Equilibrium Model:}\label{subsec:equilibrium}

We present here a novel galaxy evolution model known as
an equilibrium model, based on a simple set of equations that we
have shown to well-approximate galaxy evolution in full hydrodynamic
simulations \citep{2008MNRAS.385.2181F}.  The equilibrium models
make the ansatz that galaxies grow along a slowly-evolving equilibrium
between accretion, feedback, and star formation described by
\citep{2008MNRAS.385.2181F,2010ApJ...718.1001B,dav12,2013ApJ...772..119L,2014MNRAS.444.2071D}
\begin{equation}\label{eqn:equil}
\dot{M}_{\rm in} = \dot{M}_{\rm out} + \dot{M}_{\rm *},
\end{equation}
where the terms are the mass inflow rate into the ISM, mass
outflow rate from the ISM, and star formation rate, respectively.  This
is essentially a mass balance equation, plus our {\it equilibrium
assumption} that the gas reservoir in galaxies is non-evolving.

Using Equation \ref{eqn:equil}, one can derive equations for the
star formation rate and heavy element content of a galaxy \citep[see][and the Appendix]{dav12}:
\begin{equation}\label{eqn:SFR}
\dot{M}_{\rm *} = \frac{\zeta\dot{M}_{\rm grav}+\dot{M}_{\rm recyc}}{1+\eta},
\end{equation}
\begin{equation}\label{eqn:Z}
Z_{\rm ISM} = \frac{y\dot{M}_{\rm *}}{\zeta\dot{M}_{\rm grav}}
\end{equation}
where $\dot{M}_{\rm grav}$ is the gravitational (baryonic) inflow of dark 
matter halos obtainable from robust dark matter-only simulations
\citep{2009Natur.457..451D}, $y$ is the metal yield derived from
models of stellar nucleosynthesis \citep{2009ARA&A..47..481A}, 
and $\dot{M}_{\rm recyc}$ is the accretion rate of material that
was previously ejected.  

Here we consider metallicities in units of solar, and we assume
that the yield is exactly solar metallicity; all metallicities would
scale directly with variations in this assumption.  Metals here are
assumed to arise from core-collapse supernovae and reside in ISM
gas, and hence are most closely related to observations of e.g. the
gas-phase oxygen abundance.  We further note that this model is
most appropriately applied to {\it central} galaxies, as satellites
are subject to environmentally-dependent processes that we will not
consider here.  In principle, satellites could be evolved within
an equilibrium model by including environmentally-dependent quenching
processes, but we leave that for future work and here focus on
central galaxies only, for which we assume no explicit environmental
dependence.

The equilibrium relations can be specified using three unknown
variables that encapsulate some of the most poorly understood aspects
in galaxy formation models:
\begin{itemize}
\item $\eta$, the mass loading factor, which is the mass ejection rate 
from the ISM in units of the galaxy's star formation rate; 
\item $\zeta$, the preventive feedback parameter, which quantifies how 
much of the gas entering the galaxy's halo is prevented from reaching the ISM; 
\item $t_{\rm rec}$, the wind recycling time, which quantifies the typical 
timescale for recycling previously ejected gas back into the ISM.  
\end{itemize}
These ejective, preventive, and recycling parameters describe how
feedback processes move baryons into and out of galaxies, and hence
are referred to as {\it baryon cycling parameters}.  We note that
\citet{dav12} used a parameter defined by the enrichment of infalling
gas to quantify wind recycling, but current models have tended to
focus on $t_{\rm rec}$~\citep[e.g.][]{2013MNRAS.431.3373H}, hence we regarded
it to be interesting for our model to constrain this parameter directly.

\subsection{Parameterization}\label{subsec:parameterization}

\begin{figure}
  \includegraphics[height=0.45\textwidth, angle=270]{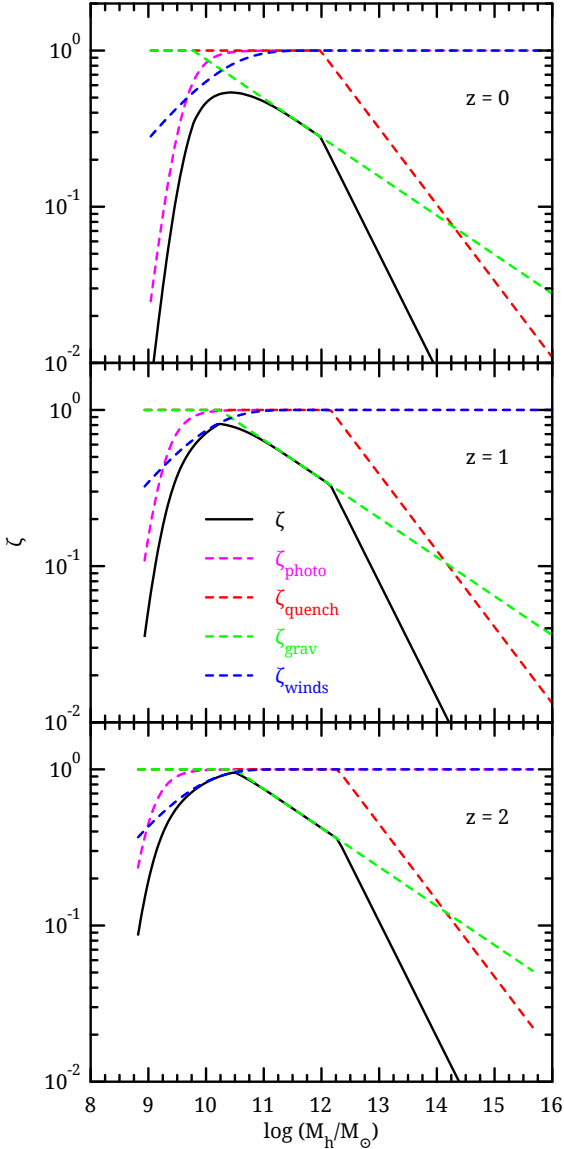}
  \caption{Preventive feedback parameter $\zeta$ at $z=0,1,2$,
showing the contributions from each of the terms in Equation~\ref{eqn:zeta}.
This is obtained for our best-fit model described in \S\ref{subsec:MCMC}.
}
\label{fig:zeta}
\end{figure}

The goal of the equilibrium model is to constrain the poorly-known
baryon cycling parameters using observations.  With little prior
information on their values and dependences, we assume they have
simple dependences on mass and redshift, as follows:
\begin{equation}\label{eqn:etaparam}
\eta = \left(\frac{M_h}{10^{\eta_1+\eta_2\sqrt{z}}}\right)^{\eta_3}
\end{equation}
\begin{equation}\label{eqn:trecparam}
t_{\rm rec} = \tau_1\times10^9{\rm yr}\times(1+z)^{\tau_2} \left(\frac{M_h}{10^{12}}\right)^{\tau_3}
\end{equation}
\begin{equation}\label{eqn:zeta}
\zeta = \zeta_{\rm photo}\times\zeta_{\rm winds}\times\zeta_{\rm grav}\times\zeta_{\rm quench}.
\end{equation}
$\zeta_{\rm photo}$ reflects prevention of infall owing to metagalactic
photo-ionisation, and is parameterised based on the filtering mass
from \citet{2008MNRAS.390..920O} as described in \citet{dav12}.
$\zeta_{\rm winds}$ measures the prevention of gas inflow owing to
energy input into circum-galactic gas by outflows
\citep{2010MNRAS.406.2325O,2013MNRAS.436.2892N}, and is parameterised
as in \citet{dav12}.  These two quantities only impact fairly
low-mass halos, so our results here are not very sensitive to these
choices.  Meanwhile, $\zeta_{\rm grav}$ suppresses inflow in high-mass
halos by the gas heating due to gravitational structure formation,
which is well-constrained from hydrodynamic simulations
\citep{2011MNRAS.417.2982F,dav12}.  Finally, $\zeta_{\rm quench}$
describes heating from physical processes that suppress star formation
in massive halos such as black hole accretion
\citep{2008MNRAS.391..481S,2015MNRAS.447..374G}.  The quenching
mass significantly impacts our results, and hence we will leave
this as a free parameter, parameterized as
\begin{equation}\label{eqn:zetaparam}
 \zeta_{\rm quench} = {\rm MIN}\left[1,\left(\frac{M_h}{M_q}\right)^{\zeta_1}\right], 
 \frac{M_q}{10^{12} {\rm M}_\odot} = (0.96 + \zeta_2 z).
\end{equation}
We further make an ansatz that recycling shuts off at $M_h>M_q$,
which is physically motivated since quenching should prevent any
re-accretion.  The functional form of $\zeta(M_h)$ at $z=0,1,2$
is shown in Figure~\ref{fig:zeta}, broken down by the contributions from
the various terms, with $\zeta_{\rm quench}$ taken for our best-fit 
fiducial model described in \S\ref{subsec:MCMC}.

The parameterizations of $\eta$, $\zeta_{\rm quench}$, and $t_{\rm
rec}$ are mostly chosen to have generic dependences on halo mass
and redshift.  One choice that is not intuitive is where we utilize
$\sqrt{z}$ in the $\eta$ parameterization.  This was done because
using just $z$ resulted in overly rapid evolution at $z\gg 2$ when
constrained to fit data at $z\leq 2$, whereas our chosen parameterization
mitigates this. Another choice is the value of $M_q=0.96\times
10^{12} {\rm M}_\odot$ at $z=0$, which we took from \citet{2012MNRAS.427.1816G}
as the typical halo mass where their hot gas halo quenching kicked
in; had we chose a more ``canonical" value of $10^{12}{\rm M}_\odot$ the
results would be indistinguishably different.  There are undoubtedly
other variations of parameterization that could have been chosen,
but we will show that these prove to be a viable descriptor.

Hence we have eight free parameters that describe baryon cycling in the
equilibrium model framework.  Our goal is thus to constrain these
parameters, thus providing insights into the physical mechanisms
that govern baryon cycling.

\subsection{MCMC and Datasets}\label{subsec:MCMC}

In order to constrain the baryon cycling parameters, we employ a
Bayesian Monte Carlo Markov Chain (MCMC) approach to simultaneously
constrain all the free parameters against key observed galaxy scaling
relations.  We use the publicly available CosmoMC
\citep{2002PhRvD..66j3511L} code, over the entire parameter space
of $\{\eta_1,\eta_2,\eta_3,\tau_1,\tau_2,\tau_3,\zeta_1,\zeta_2\}$.
For constraints, the observations we employ are recent measurements
of two well-known galaxy scaling relations that relate the halo
mass, stellar mass, and metallicity of galaxies at $z=0,1,2$:  

\begin{itemize}
\item The $M_*-M_h$ (SMHM) relation, for which we average two recent
determinations from abundance matching
\citep{2013ApJ...770...57B,2013MNRAS.428.3121M}, combining their
uncertainties in quadrature.  We note that, given a halo mass function
from $\Lambda$CDM and the assumption of abundance matching, this
is equivalent to constraining to the stellar mass function.

\item $M_*-Z$ relation (MZR), for which we use $z=0$ data as determined
from direct abundance measures in stacked SDSS spectra
\citep{2013ApJ...765..140A}, and for $z=1$, we use the data from
\cite{2014ApJ...791..130Z} with $0.1$ dex uncertainties and rescaling
the saturation metallicity to be the same as that in
\cite{2013ApJ...765..140A} ($\approx8.69$, which is comparable to
the solar metallicity value).  The rescaling is justified owing to
the large variations in absolute calibrations among different
metallicity indicators~\citep{2008ApJ...681.1183K}.  At $z\approx 2$, we
employ measurements from both Keck Baryonic Structure Survey (KBSS)
\citep{2014ApJ...795..165S} and MOSFIRE Deep Evolution Field (MOSDEF)
survey \citep{2015ApJ...799..138S}, but these data sets' calibrations
are even less certain because the physical conditions in the ISM
at those epochs may be substantially different than in today's
galaxies where those indicators are calibrated.  To account for
this, we take the average of the observed binned mass-metallicity
data sets for their two quoted calibrations (N2 and N2O2), with
errors enclosing the total uncertainties among both calibrations.

\item The $M_*-$SFR relation, often called the star-forming galaxy
main sequence~\citep[MS;][]{2007ApJ...660L..43N}.  We compare to observations
from \cite{2014ApJS..214...15S} at $z=0$, and a combination of
\cite{2014ApJ...795..104W}(blue points) and \cite{2015A&A...575A..74S}
(green points) data sets, adding an additional $0.14$ dex in
systematic uncertainties computed by the variance between the two
data sets, at $z=1,2$.  All these data sets are for {\it star-forming}
galaxies only, hence we do not use any data points beyond the
quenching mass $M_q$ at that redshift, i.e. we restrict the fit to
the range $10^9 {\rm M}_\odot\lesssim M_*\lesssim M_q$.  However, for
comparison we still show the data and model beyond $M_q$. For further
comparison, we also show the measurements of stellar mass--SFR
relation by \cite{2007ApJS..173..267S} for $z=0$ and
\cite{2014ApJS..214...15S} for $z=1,2$.

\end{itemize}

We constrain our model parameters by maximizing the likelihood
function $L \propto \exp (-{\cal L})$, where ${\cal L}$ is estimated
from
\begin{equation}
 {\cal L} = \frac{\chi^2}{2} = \frac{1}{2}\sum_{\alpha=1}^{n_{\rm obs}} \left[
\frac{\Gamma_{\alpha}^{\rm obs} - \Gamma_{\alpha}^{\rm th}}{\sigma_{\alpha}}
\right]^2
\end{equation}
where  $\Gamma_{\alpha}$ represents the set of $n_{\rm obs}$ data
used in this work and $\sigma_{\alpha}$ are the corresponding
error-bars.  We then run a number of separate MCMC chains to maximize
$L$ until the usual Gelman and Rubin convergence criterion
\citep{An98stephenbrooks} is fulfilled.

\section{MCMC constraints: Best-fit values}\label{sec:fits}

\begin{table}
\begin{center}
\begin{tabular}{l|c}
Parameters & Best-fit value and 2-$\sigma$ errors\\
\hline
\hline\\
$\eta_1$ & ~$10.98_{-0.20}^{+0.11}$ \\\\
$\eta_2$ & ~~$0.62_{-0.11}^{+0.13}$ \\\\
$\eta_3$ & $-1.16_{-0.13}^{+0.15}$ \\\\
$\tau_1$ & ~~$0.52_{-0.11}^{+0.50}$ \\\\
$\tau_2$ & $-0.32_{-0.41}^{+0.13}$ \\\\
$\tau_3$ & $-0.45_{-0.14}^{+0.25}$ \\\\
$\zeta_1$ & $-0.49_{-0.19}^{+0.18}$  \\\\
$\zeta_2$ & ~~$0.48_{-0.26}^{+0.28}$ \\\\
\hline
\end{tabular}
\caption{MCMC results: best-fit values and 95\% confidence limits on the all eight parameters.} 
\label{tab:MCMC-results}
\end{center}
\end{table}

\begin{figure*}
  \includegraphics[height=0.9\textwidth, angle=270]{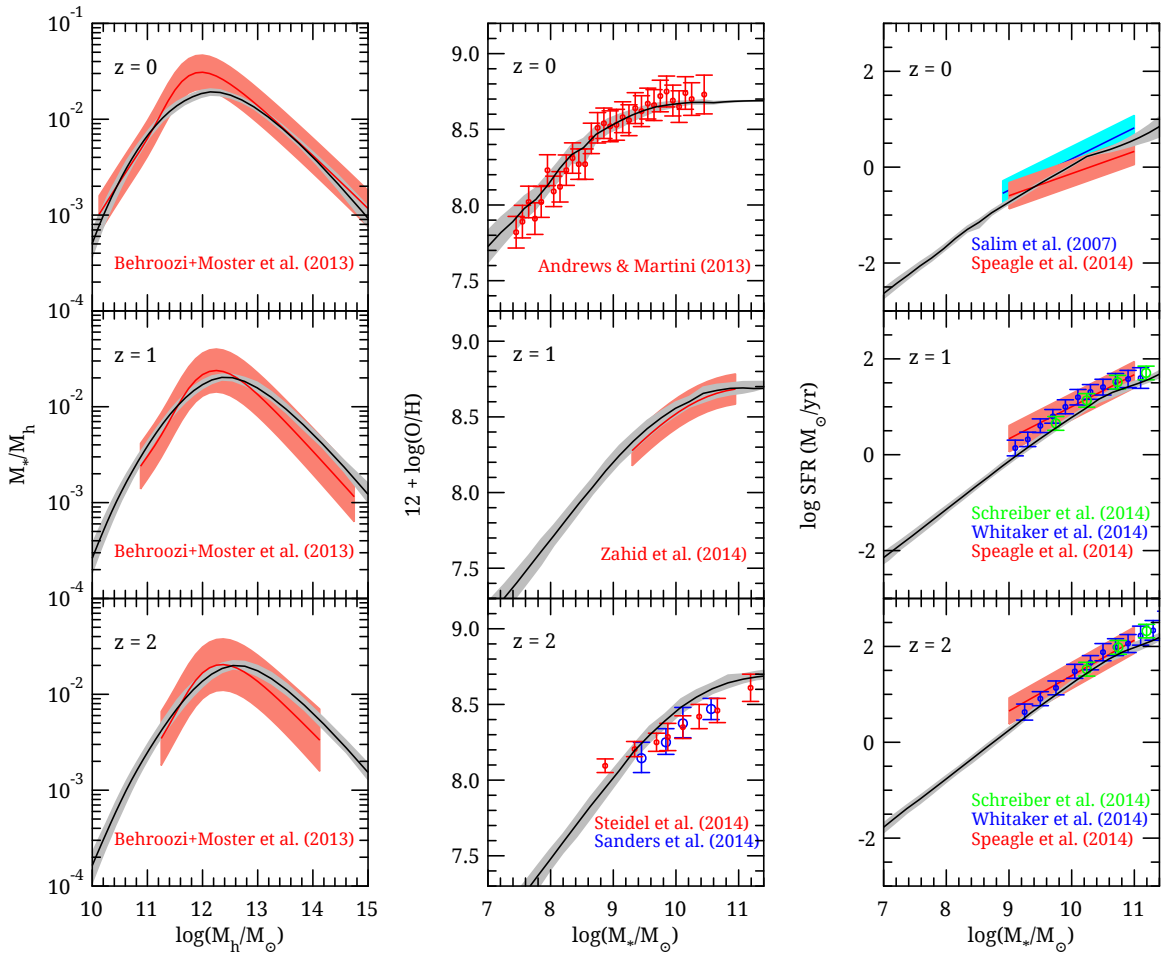}
  \caption{
The marginalized posteriori distribution predicted from MCMC constraints 
on our equilibrium model for galaxy scaling relations at $z=0,1,2$. 
Left panel: The stellar mass--halo mass (SMHM) relation; 
middle panel: mass--metallicity relation (MZR); and 
right panel: the SFR--$M_*$ relation (MS). 
The solid black lines denote the best-fit model 
and the gray shaded regions correspond to their 2-$\sigma$ confidence
limits for all panels.
The observational determinations utilized as MCMC constraints are shown as 
the red line with $1\sigma$ uncertainties (in the case of published
fitting relations) or data points with $1\sigma$ errors (in the case of 
direct determinations).  The specific observations employed are 
described in \S\ref{subsec:MCMC}.}
\label{fig:mcmc_3data}
\end{figure*}

Using CosmoMC, we obtain statistically robust constraints on the
baryon cycling parameters, as listed in Table \ref{tab:MCMC-results}.
The best-fit model is shown in Figure \ref{fig:mcmc_3data}, compared
to our set of constraining data at $z=0,1,2$ in the two left
panels.  The use of MCMC mitigates the issue of solutions being
caught in local minima.  The overall agreement is quite good, with
a reduced chi-squared value of $\chi^2_\nu\approx 1.64$.  We use
the Bayesian evidence to demonstrate that removing any one of these
parameters is not statistically favored (see \S\ref{sec:evidence}).

In detail, there are small systematic discrepancies among the various
data.  In the $M_*-M_h$ relation, our model does not reproduce the
highly peaked behavior seen at $M_h\sim 10^{12}{\rm M}_\odot$ 
\citep[also reported by][]{2013MNRAS.431.3373H,2015MNRAS.446..521S}, which
corresponds to the sharpness of the cutoff in the stellar mass
function above $L^*$.  This peakiness is driven by the
\citet{2013ApJ...770...57B} determination, whereas the
\citet{2013MNRAS.428.3121M} determination is in better agreement
with our predictions. 
We have also checked that, when constraining the model against $z=0$ 
data alone, it can reproduce the observed peak in SMHM quite accurately. 
But when we try to match the galaxy properties for all epochs up to $z=2$,
the best-fit MCMC results underpredicts this peak at $z=0$. Given the 
increasing uncertainties at high redshifts, this might be avoided by 
putting some extra weight on the $z=0$ constraints in the MCMC run. 
Again the equilibrium model does not account for galaxy mergers, while 
they can contribute (perhaps little) to the total stellar mass of a galaxy. 
Then the added stellar content can help to match the observed peak better. 
In our future work we will try to incorporate such effects.
Overall, the $M_*-M_h$
relation shows a faint-end slope and bright-end cutoff that well
matches data, at all redshifts shown.

The mass-metallicity relation also shows good agreement, at least
at $z=0,1$.  At $z=2$, it appears the observations are slightly
below the inferred value, but given the calibration uncertainties
we do not take this discrepancy too seriously.  A key trend that
is predicted by our model is that the turnover $M_*$ increases with
redshift while the faint-end slope remains similar, which is in
agreement with that inferred from observations by
\cite{2014ApJ...791..130Z}.  More broadly, the simultaneous agreement
of the faint-end slope of the $M_*-M_h$ relation and the mass-metallicity
relation is non-trivial.  Simulations that match the mass-metallicity
faint end tend to have too shallow a faint end for $M_*-M_h$ \citep[or
equivalently, too steep a faint end for the stellar mass
function;][]{dav11a}.  Increasing outflows in low-mass galaxies
to better match $M_*-M_h$ results in too steep a mass-metallicity
relation~\citep{dav13}.  Avenues to mitigate this include some
additional preventive feedback not currently arising in
simulations~\citep{2014MNRAS.438.1985T}, or a lower efficiency of converting
gas into stars at low masses ~\citep{2007ApJ...655L..17B,2010ApJ...714..287G}.

The predicted MS evolves upwards in SFR at a given $M_*$ with a
mostly invariant slope, in general agreement with observations.
The amplitude evolution is broadly consistent with observations,
though it tends to be somewhat low compared to observations at
high-$z$ (as also seen in~\citealt{2013ApJ...769..148H}),
by no more than a factor of 2.  We will show in
\S\ref{subsec:datarequired} that a better match can be achieved to
the MS, at the expense of a poorer match to the MZR, which may still
be acceptable given the more substantial systematic uncertainties
on metallicity measures in higher-$z$ galaxies.  The amplitude
evolution of the MS out to $z\sim 2$ has proved particularly difficult
for both semi-analytic models and simulations to match, generally
showing discrepancies at the factor of $2-4$ level at $z\sim 2$
when constrained (or predicted) to match at $z\sim 0$
\citep{2008MNRAS.385..147D,2015MNRAS.447.3548S}.  Hence the level
of agreement achieved by the equilibrium model from $z\sim 0-2$ is
comparable to or better than in SAMs or simulations, which is a
noteworthy success given its simplicity.  We explore MS evolution
to higher redshifts in \S\ref{subsec:hiz}.

\section{Assessing the MCMC fit}\label{sec:Bayesian}

\subsection{Are all the parameters necessary?}\label{sec:evidence}

\begin{table}
\begin{tabular}{l|ccc}
Model & $\ln {\cal Z}$ & ${\cal K}$ & $\chi^2/{\rm DOF}$ \\ 
\hline
\hline \\
{\it True} & $-15.54\pm0.15$ &  & $1.64$ \\\\
{\it Null} ($\eta_1=12$) & $-29.42\pm0.16$ & $>10^6$ & $6.82$ \\\\
{\it Null} ($\eta_2=0$) & $-22.31\pm0.15$ & $>10^2$ & $5.23$ \\\\
{\it Null} ($\eta_3=-1$) & $-17.06\pm0.13$ & $5$ & $2.09$ \\\\
{\it Null} ($\tau_1=t_{\rm H}$) & $-18.63\pm0.14$ & $22$ & $3.11$ \\\\
{\it Null} ($\tau_2=0$) & $-17.97\pm0.13$ & $11$ & $2.59$ \\\\
{\it Null} ($\tau_3=0$) & $-18.93\pm0.15$ & $30$ & $3.48$ \\\\
{\it Null} ($\zeta_1=0$) & $-27.31\pm0.14$ & $>10^5$ & $6.17$ \\\\
{\it Null} ($\zeta_2=0$) & $-19.84\pm0.15$ & $74$ & $4.03$ \\\\
\hline
\end{tabular}
\caption{MultiNest results: the mean and standard deviation of the
Bayesian evidence and corresponding $\chi^2$ values for the {\it
true} model with eight parameters and the {\it null} hypotheses with seven
parameters.  The Bayes factor ${\cal K}$ is the ratio of the evidences
between the {\it true model} and the {\it null model}.  It is $>1$
in all cases, formally indicating the preference for the our
eight-parameter model, although the inclusion of $\eta_3$ and $\tau_2$
is only weakly preferred. Note that, the baryon cycling parameters
for {\it true model} are constrained to have best-fit values of
$\eta = \left(M_h/10^{10.98+0.62\sqrt{z}}\right)^{-1.16}$; $t_{\rm
rec} = 0.52\times10^9{\rm yr}\times(1+z)^{-0.32}
\left(M_h/10^{12}\right)^{-0.45}$; and $\zeta_{\rm quench} =
\left[M_h/(0.96\times10^{12}+0.48z\times10^{12})\right]^{-0.49}$}
\label{tab:multinest-results}
\end{table}

As described in \S\ref{subsec:MCMC}, we have chosen to represent
our baryon cycling parameters with eight free parameters.  The resulting
reduced $\chi^2\sim 1.6$ shows that these parameters are sufficient
to achieve a good statistical match.  A natural question then to
ask is whether all of these eight parameters are formally necessary.
We use the Bayesian evidence to assess this.

For a given model $M$ with the data ${\bf D}$ and a set of parameters ${\bf \Theta}$,
one can estimate the posterior probability density of the model parameters
from the Bayes' theorem as~\citep{2008MNRAS.384..449F,2013arXiv1306.2144F}
\begin{equation}\label{eqn:bayes}
 {\rm Pr}({\bf \Theta}|{\bf D},M)=\frac{{\rm Pr}({\bf D}|{\bf \Theta},M){\rm Pr}({\bf \Theta}|M)}{{\rm Pr}({\bf D}|M)}
\end{equation}
where, ${\rm Pr}({\bf D}|{\bf \Theta},M)\equiv L({\bf \Theta})$ is the likelihood of
the data, ${\rm Pr}({\bf \Theta}|M)\equiv\pi({\bf \Theta})$ is the prior, and
${\rm Pr}({\bf D}|M)\equiv{\cal Z}$ is the Bayesian evidence.
The evidence is then determined by calculating the average of the likelihood
over the prior for that model
\citep{2008MNRAS.384..449F,2009MNRAS.398.1601F}:
\begin{equation}\label{eqn:evidence}
 {\cal Z} = \int L({\bf \Theta})\pi({\bf \Theta}){\rm d}^D{\bf \Theta}
\end{equation}
where, $D$ is the dimension of the parameter space. The
evidence penalized models with more parameters via an increased 
volume of the likelihood posterior, while it penalizes models
that fit poorly by a lowered likelihood.  Assuming
there is no strong a priori reason for favoring one model over
another, the selection between two models can thus be assessed by 
comparing their evidences \citep{2009arXiv0911.1777W,2013arXiv1306.2144F}.
If the evidence ${\cal Z}_1$ for model $M_1$ is larger than the evidence
${\cal Z}_2$ of the model $M_2$, we can say the former model is statistically
more favored than the later.  This can be quantified by 
the Bayes factor ${\cal K}={\cal Z}_1/{\cal Z}_2$; ${\cal K}>1$
means that $M_1$ is more preferable by the data under consideration
than $M_2$.

We define a {\it null model} by fixing any single one of the eight
parameters as follows: $\{\eta_1=12, \eta_2=0, \eta_3=-1, \tau_1=t_{\rm H},
\tau_2=0, \tau_3=0, \zeta_1=0, \zeta_2=0\}$, where $t_{\rm H}$ is the
Hubble time i.e. the age of the Universe at a given redshift.  We
then vary all other seven parameters to obtain a best fit, 
and calculate the Bayesian evidence using the MultiNest
\citep{2008MNRAS.384..449F,2009MNRAS.398.1601F} code.  We then
compare this to the evidence using all eight parameters to decide
whether this given parameter is necessary, or whether it can simply
be set to a ``natural" value.  In most cases, we take the natural
value to be the one that removes the dependence on the quantity
(e.g. redshift or halo mass), though in several cases we choose
a value that is canonically preferred.

In Table \ref{tab:multinest-results}, we compare the evidences
between our current model with eight parameters ({\it true model})
and the eight {\it null models}. Here, the Bayes factor ${\cal K}$
is the ratio of the evidences between the {\it true model} and the
{\it null model}.  We also show the corresponding $\chi^2$-values
over the degrees of freedom (${\rm DOF}$).  The ``true" (eight-parameter)
model has a reduced $\chi^2=1.64$, while the null models have a
higher value.  But it is still necessary to calculate the evidence
to determine if the lower $\chi^2$ in the true model is outweighed by
increased freedom of having more parameters.

Table \ref{tab:multinest-results} shows that the evidence is always
larger for the {\it true} model than any of the {\it null} hypotheses.
This suggests that all eight parameters are formally necessary in
order to better constrain the galaxy formation model.  Given the
acceptable value of the reduced $\chi^2\approx 1.6$, this further
suggests that these eight parameters are sufficient to describe the
fitted observations.
The Bayes factor ${\cal K}$ is very large for some cases like the
quenching feedback parameters ($\zeta_1$, $\zeta_2$) or the 
normalization of the mass loading factor ($\eta_1$, $\eta_2$),
suggesting that the inclusion of those parameters is strongly favored
for obtaining a good fit. On the other hand, one can see that the inclusion of
other parameters e.g. the recycling parameters is only weakly preferred. 
It is also somehow evident from the fact that, the majority of our 
observational constraints are at high masses above where the quenching
effect is significant, whereas our model shuts off the recycling at
that mass scale.

\subsection{Are all the parameters independent?}

\begin{figure*}
   \includegraphics[height=0.9\textwidth, angle=270]{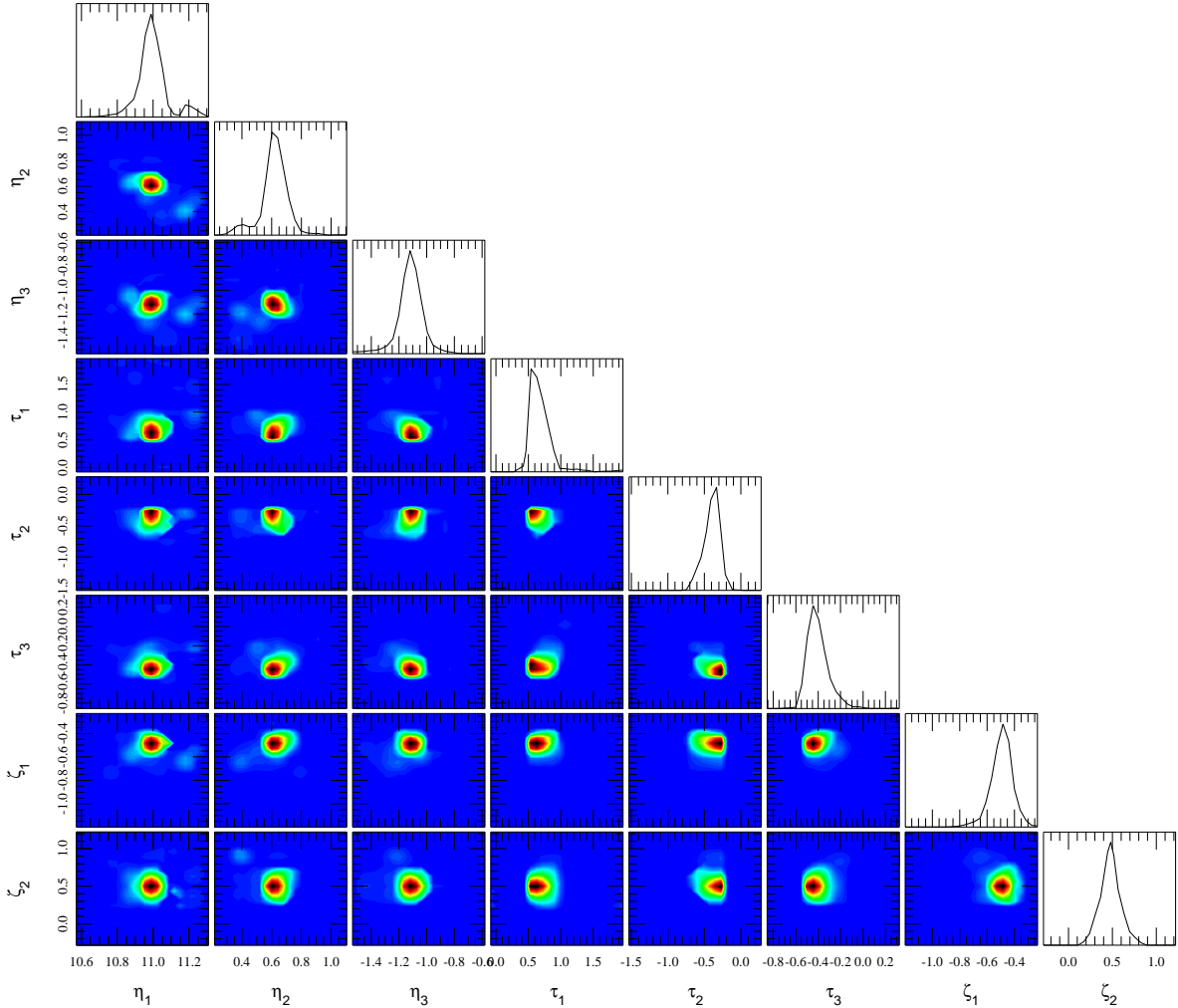}
\caption{Marginalized 2-D contours for all eight parameters
from MCMC analysis. Panels along the diagonal show the 1-D marginalized
probability distributions for each parameter, while the off-diagonal panels
indicate the joint probability distributions of them.}
\label{fig:contours}
\end{figure*}

The Bayesian MCMC approach has the advantage of being able to view
the posterior probability for each parameter, and particularly to
determine whether parameters are significantly correlated.  This
offers another way to ensure the parameterizations we have adopted
are necessary and sufficient.

Figure \ref{fig:contours} shows the 1-D (diagonal panels) and 2-D
(off-diagonal panels) marginalized posterior probability distributions
for all eight parameters.  The generally circular appearance of the
best-fit covariances indicates that none of parameters are significantly
correlated with each other.  The concentrated peak in each panel
shows that, while other solutions exist, the best-fit solution is
strongly preferred.  These panels also shows the range considered
for each variable, over which we have assumed a flat prior; the
best-fit values are comfortably within the assumed range.  This
demonstrates that the parameterizations as well as the range
for the prior that we have taken are reasonable choices.

\subsection{Are all the data required?}\label{subsec:datarequired}

\begin{figure*}
  \includegraphics[height=0.9\textwidth, angle=270]{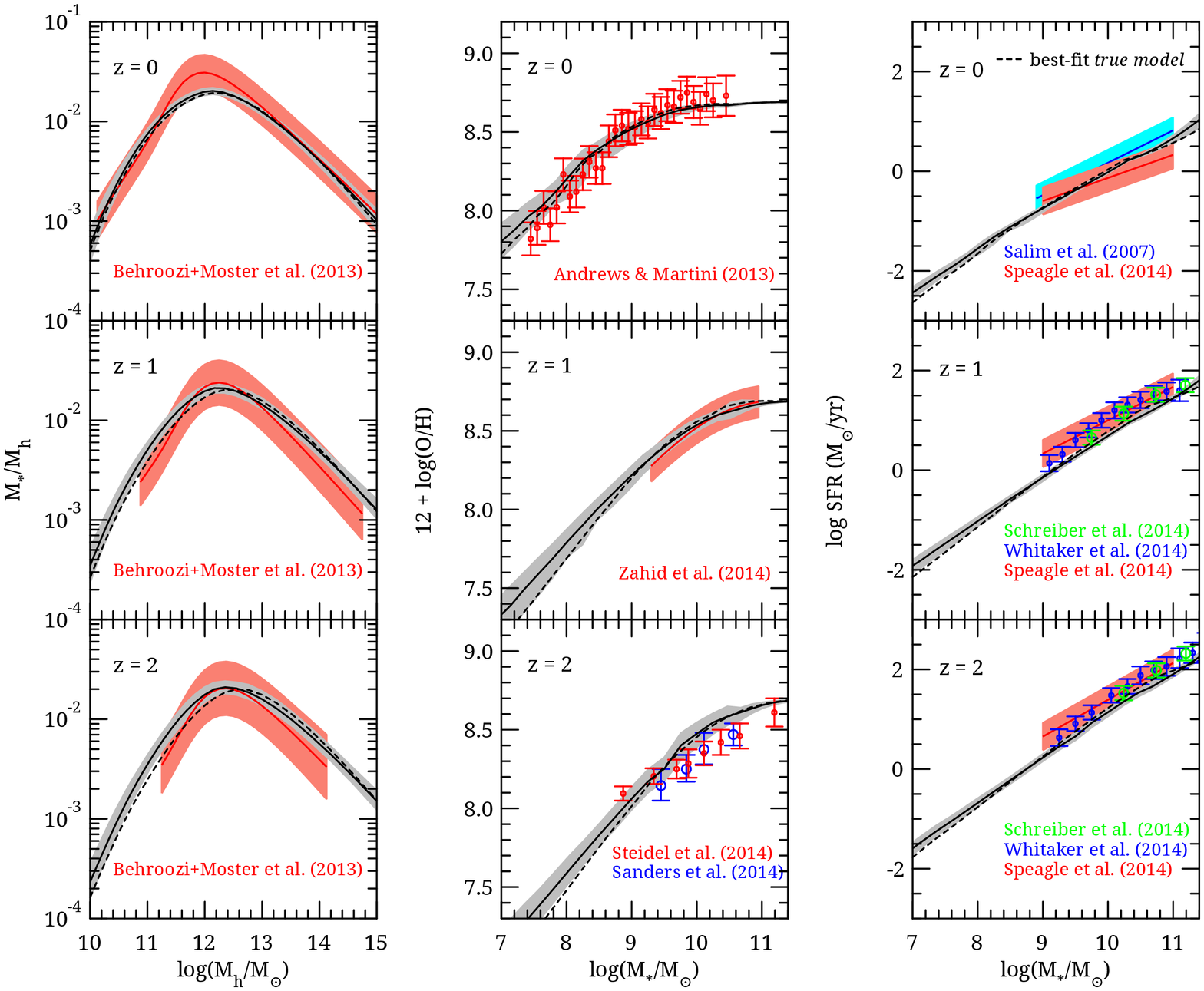}
\caption{
The marginalized posteriori distribution from MCMC analysis for 
our considered scaling relations at $z=0,1,2$, here fitting only to two
relations, namely SMHM and the MZR.  Data are shown as described
in Figure~\ref{fig:mcmc_3data}. The best-fit {\it true} model with all three data sets
from Figure~\ref{fig:mcmc_3data} is also plotted here as short-dashed lines for comparison.}
\label{fig:mcmc_data12}
\end{figure*}

\begin{figure*}
  \includegraphics[height=0.9\textwidth, angle=270]{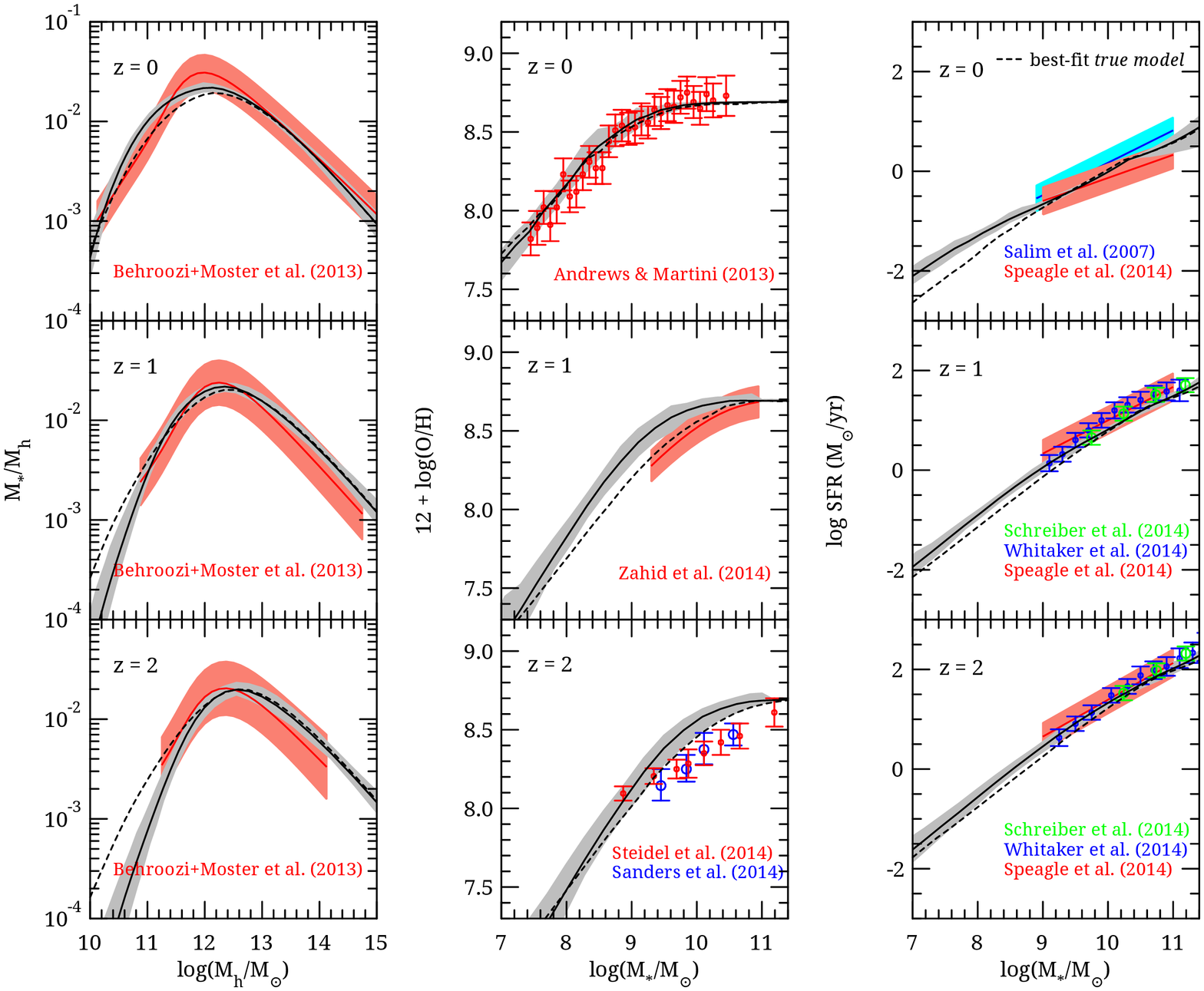}
\caption{
The marginalized posteriori distribution from MCMC analysis for 
our considered scaling relations at $z=0,1,2$, here fitting only to 
SMHM and the MS.  Data are shown as described
in Figure~\ref{fig:mcmc_3data}.}
\label{fig:mcmc_data13}
\end{figure*}

\begin{figure*}
  \includegraphics[height=0.9\textwidth, angle=270]{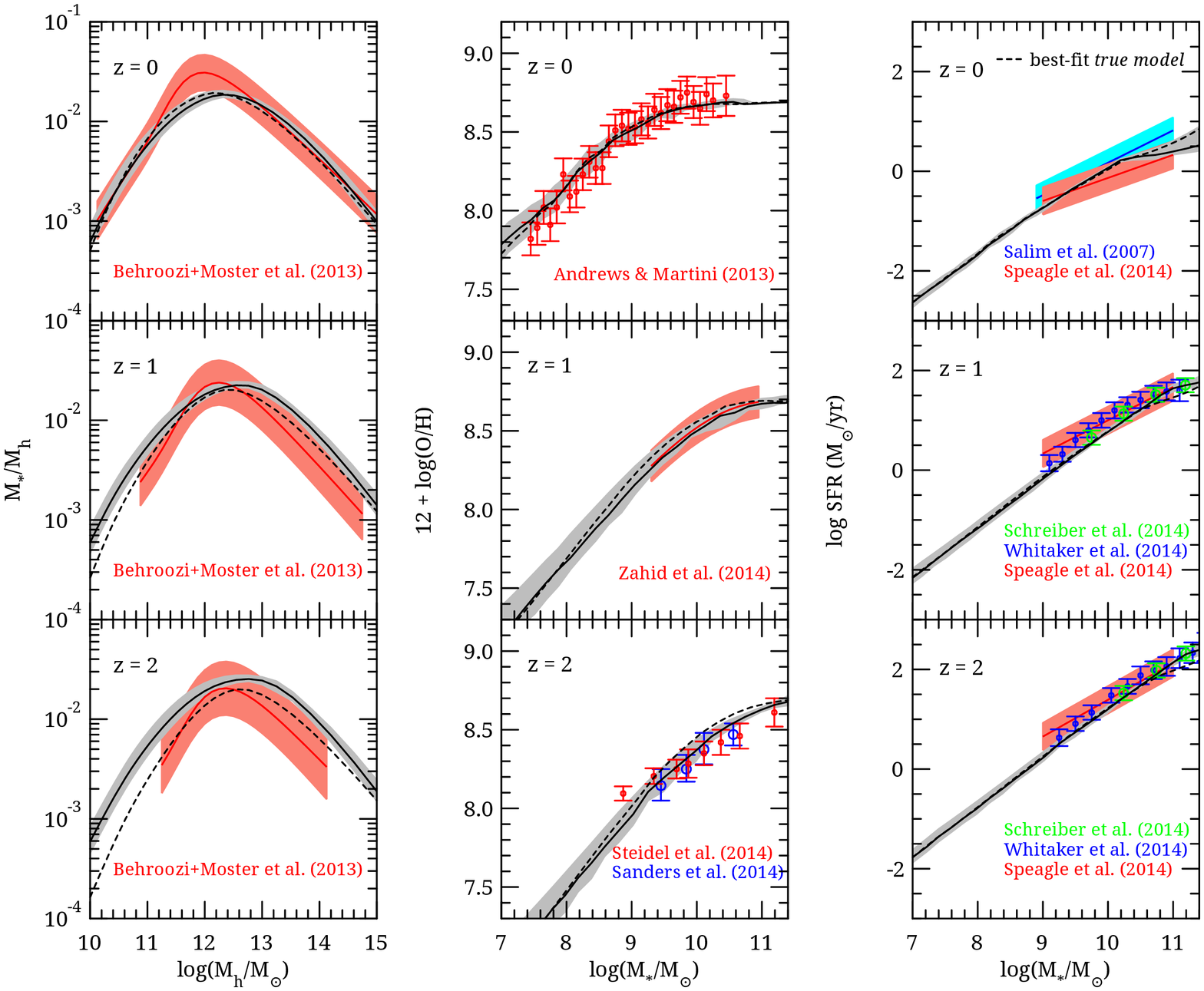}
\caption{
The marginalized posteriori distribution from MCMC analysis for 
our considered scaling relations at $z=0,1,2$, here fitting only to 
the MZR and the MS.  Data are shown as described
in Figure~\ref{fig:mcmc_3data}.}
\label{fig:mcmc_data23}
\end{figure*}

Here we examine whether all 3 scaling relations we have considered
are necessary to obtain a good fit.  The way we assess this is
to fit only two of the three relations, and determine how well
the resulting best-fit model {\it predicts} the third.

Figure \ref{fig:mcmc_data12} shows the result of fitting only to
the SMHM and MZR relations, and predicting the MS.  The solid line
is the new best-fit, while the dashed line shows the original model
fit to all three data sets reproduced from Figure \ref{fig:mcmc_3data}
for comparison.  The overall fit remains mostly unchanged.  The
agreement is slightly worse for the MS, as might be expected since
we no longer utilizing that data as a constraint.  However, the fit
is marginally better for other data sets.  In the end, $\chi^2_\nu$
for this model fit is 1.51 just considering the fit to the SMHM and
MZR.  The overall $\chi^2_\nu$ for all three data sets is 1.68,
which is slightly worse, but still within acceptable uncertainties.
This demonstrates the rather remarkable result that even just fitting
to two of these scaling relations (SMHM and MZR), the equilibrium
model is able to {\it predict} the independently-determined MS to
good accuracy.

Now, one could argue that constraining to the stellar mass growth
as a function of halo mass out to $z=2$ naturally assures that the
SFR will be reproduced, assuming that halo mass growth is properly
tracked. Also note that, for our model, matching the MZR for a given $M_*$
essentially forces to match the SFR via Equation~\ref{eqn:Z}.
While true in principle, current models that
broadly match the SMHM relation and its evolution still
fail to match the MS evolution at the $\ga\times 2$
level~\citep{2012MNRAS.426.2797W}.  Hence
the agreement predicted by the equilibrium model to better than a 
factor of 2 from $z=0-2$ is non-trivial.  

To more fully explore our model, we can fit to the SMHM relation and the
MS, and then predict the MZR.  In this case, the metallicity is not
being used as a constraint anywhere, so we can assess whether the
metallicity evolution is predicted correctly in this model. Figure
\ref{fig:mcmc_data13} shows the result.  The fit to the MZR is still
excellent at $z=0$, and is not much different at $z=2$, but tends
to overshoot the $z=1$ data somewhat by having too low a turnover
mass.  Hence it is not quite as good as before, but given the systematic
uncertainties in metallicity measures~\citep{2008ApJ...681.1183K} it is
likely still acceptable. The general trend of an MZR with an
upwardly-evolving MZR turnover mass remain the same, which shows
that this is a generic prediction for an equilibrium model with
reasonable constraints. So we find that, this model can match
the SFR data more accurately at least for low-mass galaxies, but overpredicts
the MZR owing to Equation~\ref{eqn:Z} (a similar trend is also seen
in ~\citealt{2013ApJ...769..148H}). This could happen due to the
yield value assumed here and can be avoided by taking a smaller $y$ or
making it as a free parameter (effectively equivalent to 
varying the IMF;~\citealt{2013MNRAS.436.2892N}) which will be 
incorporated in the future work.
We further note in this case a steeper
turn-down in the SMHM relation at low masses at $z=1,2$, which
agrees better with the observationally-constrained determinations
of this.  The resulting $\chi_\nu^2$ just fitting to SMHM and MS
is 1.48, while $\chi_\nu^2$ over all data sets (including the MZR)
is now 2.01.  Hence the fit has degraded slightly overall, but
remains acceptable.

Figure~\ref{fig:mcmc_data23} lastly examines the results for the
final permutation, namely constraining to the MZR and MS and
predicting the SMHM relation.  In this case we obtain a very good
fit to the constrained relations ($\chi^2_\nu=1.55$), now predicting
the MS amplitude at $z=2$ extremely well.  However, the predicted
SMHM relation, while similar at $z=0$, increasingly overshoots the
observationally-determined values to higher redshifts, particularly
at high masses.  For this case, the overall $\chi^2_\nu=1.84$ 
when considering all three data sets.

These permutations demonstrate that the equilibrium model can fit
any two of the three scaling relations very well, and the third
relation is then also fairly well predicted.  Interestingly, the
$z=0$ values are the most closely predicted ones.  Note that we
have not constrained our model to match better at $z=0$ than at any
other redshift; all the data at all the redshifts are weighted
equivalently (i.e. by their respective uncertainties) in our MCMC
fit.  Meanwhile, fits are degraded relative to the data in our
model mostly when considering higher redshifts, where of course
systematic effects in observational determinations of these
galaxy properties likely are larger.  It will be 
interesting to see, as data at higher redshift improves, whether
observational determinations at higher redshifts become more
self-consistent between these three scaling relations within
the context of the equilibrium model, as they are at $z=0$.

Overall, the fact that the equilibrium model constrained to fit two
of the scaling relations can reasonably well predict the third
supports the idea that its underlying framework is a viable
representation for how the mean galaxy population evolves. While
one would prefer a fit of $\chi^2_\nu<1$ to claim formally perfect
agreement, the values we obtain by fitting all three or any two of
the three data sets is still remarkably good ($\chi^2_\nu\la 2$)
when considering the numerous difficult-to-quantify systematic
effects present in the data particularly to higher redshifts.

\section{Implications and Discussion}\label{sec:implications}

\subsection{Implications for baryon cycling parameters}

\begin{figure*}
 \includegraphics[height=0.6\textwidth, angle=270]{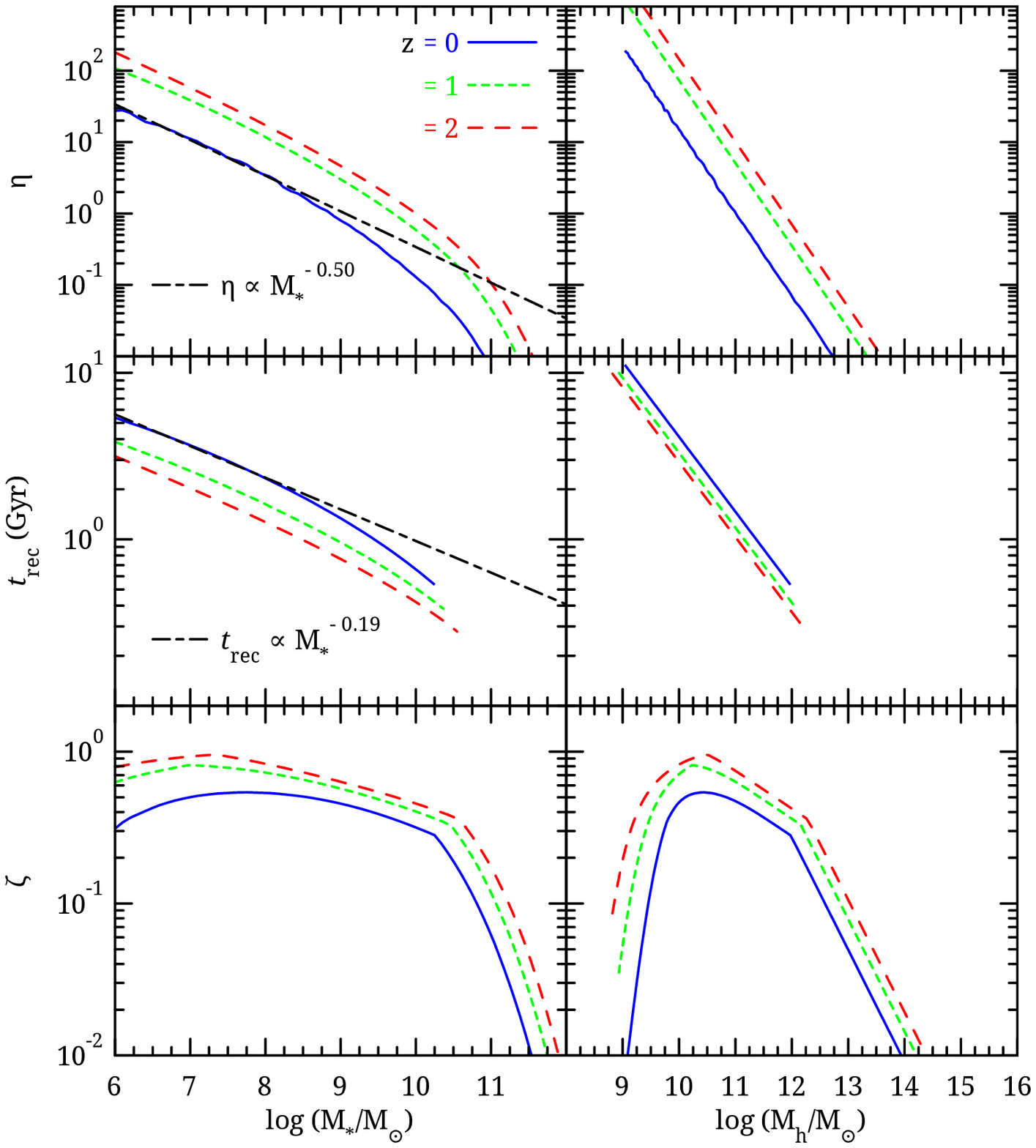}
\caption{The dependence of the baryon cycling parameters on stellar (left panel) 
and halo (right panel) masses at $z=0$ (solid blue lines), $z=1$ (short-dashed green lines) and 
$z=2$ (long-dashed red lines). For comparison, $\eta\propto M_*^{-0.50}$ and $t_{\rm rec}\propto M_*^{-0.19}$
are also plotted by short-long-dashed lines.}
\label{fig:parameter}
\end{figure*}

The crucial power of the equilibrium model is its ability to constrain
the baryon cycling parameters directly from data in a statistically
robust manner.  This then provides insights into the physics of
feedback processes, and constraints for more detailed models of
inflows and outflows.  Here we examine the predicted halo mass and
redshift dependences of the baryon cycling parameters, and briefly
discuss some physical implications.  This is intended to provide a
brief illustration of how an equilibrium model can be useful to
obtain insights into galaxy formation processes, but of course there
are many more such aspects that we will leave for future work.

Figure \ref{fig:parameter} shows the dependence of the baryon cycling
parameters on stellar and halo mass in the left panel and
right panel, respectively, at $z=0,1,2$.  While the parameters
are directly constrained based on the halo mass, we have convolved
these with our predicted $M_*-M_h$ relation to obtain the stellar
mass dependences.

The ejective feedback parameter scales as $\eta\propto M_*^{-0.5}$
at low masses, which is intermediate between scalings expected for
momentum- and energy-driven winds, suggesting that both effects are
in play \citep{2010ApJ...709..191M}.  This is broadly similar to
scalings assumed or generated in cosmological simulations that match
a wide variety of data, which have for instance assumed $\eta\propto
M_*^{-1/3}$~\citep{dav11a} to $\eta\propto v_c^{-2}\propto
M_*^{-2/3}$~\citep[roughly;][]{2014Natur.509..177V}.  It is also in
agreement with scalings predicted in ab initio simulations of
stellar-driven outflows from the Feedback in Realistic Environments
suite \citep{2013MNRAS.433.1970F,2014MNRAS.445..581H}, suggesting
that the interplay of radiation pressure and supernovae energy that
drive winds in those single-galaxy simulations is plausible. This
illustrates how the equilibrium model thus provides a way to directly
connect and constrain the predictions from high-resolution individual
galaxy simulations with galaxy population statistics across cosmic
time. Also we find that, the model robustly prefers $\eta_2$ to be
positive, which implies that galaxies are less effective at driving
outflows at high redshifts than at lower redshifts. 
Such a dependence could reflect the tendency for outflows to be less
efficient in galaxies with higher gas fractions~\citep{2015MNRAS.446.2125C}
or lower metallicities~\citep{2015MNRAS.446..521S}.

The recycling time distribution is currently much debated.  Our
model predicts that it is nearly invariant with redshift, and scales
as $t_{\rm rec}\propto M_h^{-0.5}$.  This is a significantly weaker
dependence than predicted in cosmological hydrodynamic simulations
\citep{2008MNRAS.387..577O} and inferred in SAMs
\citep{2013MNRAS.431.3373H} of $t_{\rm rec}\propto M_h^{-1}$, or
even that predicted from simple gravitational binding arguments.
Zoom simulations that employ supernova heating to drive outflows
from Christensen et al. (2015, in preparation), in contrast, predict
a similarly weak dependence of $t_{\rm rec}$, and are likewise able
to simultaneously match the $M_*-M_h$ relation~\citep{2013ApJ...766...56M}
and the MZR ~\citep{2007ApJ...655L..17B}. The interplay between outflows
and ambient circum-galactic gas remains a difficult numerical problem
for current hydrodynamical codes and depends strongly on the
particular implementation of wind driving and hydrodynamic technique. 
Hence the equilibrium model can provide important constraints on 
such difficult-to-model processes.

The quenching mass is another key parameter in galaxy formation
models.  \citet{2009Natur.457..451D} argued for a quenching mass
around $10^{12}{\rm M}_\odot$ out to $z\sim 1-2$, before which it increases
with redshift owing to cold streams penetrating hot halos.  Simulations
by \citet{2012MNRAS.427.1816G} assumed quenching happens once a hot
gaseous halo forms, and also find a quenching mass around
$10^{12}{\rm M}_\odot$ that increases mildly with redshift.  SAMs have
long implemented a halo mass above which putative AGN feedback is
able to heat gas, which again is typically $\sim 10^{12}{\rm M}_\odot$
and mostly invariant with redshift
\citep{2006MNRAS.365...11C,2006MNRAS.370..645B,2008MNRAS.391..481S}. Our model predictions
are generally in agreement, with a slowly-increasing quenching mass
with redshift.  Notably, however, our results are inconsistent with
a single invariant quenching mass at all redshifts (\S\ref{sec:evidence}).
This indicates that quenching is not purely a function of halo mass,
but depends on some other property that scales slowly with quenching
mass over time.

To further investigate why the equilibrium model robustly prefers different
scaling relations than those obtained from SAMs and hydro-simulations,
we compare the Bayesian evidence of our current {\it true} model with
the evidences from different {\it null} models. We choose the null values
for the parameters associated with $\eta$ and $t_{\rm rec}$ in a fashion that
they resemble the models with momentum-driven scaling $\eta\propto M_*^{-1/3}$
\citep{dav11a}, energy-driven scaling $\eta\propto M_*^{-2/3}$
\citep{2014Natur.509..177V} and a steeper dependence of 
$t_{\rm rec}\propto M_h^{-1}$ as inferred from SAMs~\citep{2013MNRAS.431.3373H}.
So instead of using the ``natural" null values as previously mentioned in Table
\ref{tab:multinest-results}, we now take these models as our null ones
and calculate their corresponding evidences.
We have listed our results in
Table \ref{tab:multinest-results-othernullmodels}, where one can see that
pure energy or momentum-driven outflows with no redshift dependence ($\eta_2=0$),
as assumed in those simulations, fails to give a reasonable fit to the observed data
(reduced $\chi^2\sim8$). One can get more acceptable models with those scalings by
allowing $\eta_2$ to vary (reduced $\chi^2\sim2$). In both cases, the energy-driven
winds suppress mass and metals more at low-masses, while the momentum-driven scalings
suppress less than observed. Thus an intermediate scaling ($\eta\propto M_*^{-0.5}$)
is formally more favored by the data considered here.
Table \ref{tab:multinest-results-othernullmodels} reinforces the impression from
Figure \ref{fig:contours} that observations constrain both the mass- and redshift-dependence
of galactic outflows as well as the return timescale. Judging by the Bayes factors, a
redshift evolution for $\eta$ is strongly favored, while our best-fit mass-dependence is
more weakly (but still formally) favored over canonical alternatives. Meanwhile, our
best-fits for the quenching scale and recycling are highly favored over those.
Nonetheless, the Bayes factor ${\cal K}$ is $> 1$ (also the reduced $\chi^2$ is larger)
for all null cases. This indicates that the scaling relations
for baryon cycling parameters obtained from our 8-parameter model are clearly
preferable to the parameterizations employed in those works.

\begin{table}
\begin{tabular}{l|ccc}
Model & $\ln {\cal Z}$ & ${\cal K}$ & $\chi^2/{\rm DOF}$\\ 
\hline
\hline \\
{\it True} model & $-15.54\pm0.15$ &  & 1.64\\\\
{\it Null} model with\\
$\eta\propto M_*^{-1/3}$ & $-17.30\pm0.13$ & $6$ & $2.27$\\
(varying $\eta_2$)\\\\
{\it Null} model with\\
$\eta\propto M_*^{-2/3}$ & $-17.23\pm0.13$ & $5$ & $2.18$\\
(varying $\eta_2$)\\\\
{\it Null} model with\\
$\eta\propto M_*^{-1/3}$ & $-36.16\pm0.14$ & $>10^8$ & $8.15$\\
(keeping $\eta_2=0$)\\\\
{\it Null} model with\\
$\eta\propto M_*^{-2/3}$ & $-35.80\pm0.14$ & $>10^8$ & $7.92$\\
(keeping $\eta_2=0$)\\\\
{\it Null} model with\\$t_{\rm rec}\propto M_h^{-1}$ & $-22.47\pm0.13$ & $>10^3$ & $5.28$\\\\
\hline
\end{tabular}
\caption{Comparison of the Bayesian evidences and corresponding $\chi^2$ values for our {\it true} model
with eight parameters to that obtained from different {\it null} models.
The Bayes factor ${\cal K}>1$ for all cases indicates that the {\it true} model is
more favored over any of these null models.}
\label{tab:multinest-results-othernullmodels}
\end{table}

These examples briefly illustrate how the equilibrium model can
provide insights and constraints for more physically-based models
for baryon cycling processes.  Alternatively, it can potentially
exclude models whose baryon cycling behavior is substantially
different from that predicted here, or else quantify how far off
such a model would be from matching these observed galaxy scaling
relations.  Current simulations and SAMs have, through less rigorous
parameter space exploration, arrived upon generally similar constraints
as those preferred by our model, but the equilibrium model framework
provides a more quantitative way to directly constrain baryon cycling
from observations.

\subsection{Predictions to higher redshift}\label{subsec:hiz}

So far we have seen that our model can match the observational data
sets quite reasonably from $z=0-2$.  Due to the increasingly large
observational and systematic errors, the knowledge of the galaxy
population at higher redshifts is less certain, hence we have not
included these observations as constraints on our model.  Nonetheless,
as a sanity check, we here show how our model predicts the behavior
of galaxies at earlier epochs.

\begin{figure*}
 \centerline{
   \includegraphics[height=0.9\textwidth, angle=270]{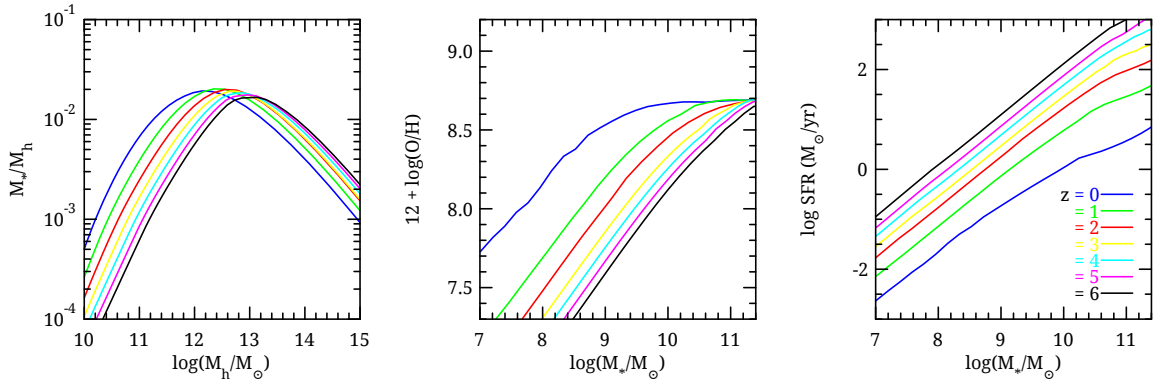}
  }
\caption{Model prediction for (i) $M_*-M_h$ relation (left panel), 
(ii) $M_*-Z$ relation (middle panel) and the $M_*$--SFR relation (right panel) 
at redshifts $z=0$ (blue), $z=1$ (green), $z=2$ (red), $z=3$ (yellow), $z=4$ (cyan), 
$z=5$ (magenta) and $z=6$ (black). Note that, the $z=0-2$ results are the same as
the best-fit MCMC analysis in Figure~\ref{fig:mcmc_3data}, whereas the higher redshifts
results are the model prediction.}
\label{fig:prediction}
\end{figure*}

In Figure \ref{fig:prediction} we show our model predictions for
the stellar mass--halo mass relation (left panel), mass--metallicity
relation (middle panel) and SFR--$M_*$ relation (right panel)
for all the redshifts up to $z=6$. Their redshift evolution
is denoted by the solid lines with different colors for different
redshifts as mentioned in the bottom of right panel.  The
$z=0-2$ results are the same as the best-fit solid lines of Figure
\ref{fig:mcmc_3data}.

The $M_*-M_h$ relation shows a clear peak in the stellar mass--halo
mass ratio which is visible at all the redshifts. The location of
the peak shifts towards higher masses with increasing redshift.
This is broadly consistent with the results out to $z\sim
2$~\citep{2012ApJ...744..159L,2013ApJ...770...57B,2013MNRAS.428.3121M}, but
at higher redshifts, the determinations diverge: \cite{2013MNRAS.428.3121M}
suggests a continued increase in the peak stellar efficiency, while
\cite{2013ApJ...770...57B} suggests a drop.  Our results are more
consistent with \cite{2013MNRAS.428.3121M}, in part because our
functional form with redshift is assumed to be monotonic.  Furthermore,
the amplitude of the peak efficiency decreases a bit at higher
redshifts, which is in agreement with recent determinations
\citep{2013ApJ...770...57B,2013MNRAS.428.3121M,2014ApJ...793...12B}. For
our best-fit model, the stellar mass--halo mass ratio peaks at
$M_h\sim10^{12}{\rm M}_{\odot}$ for $z=0$, shifting to $M_h\sim10^{13}{\rm M}_{\odot}$
for $z=6$. For very high mass galaxies, we see that the stellar
mass increases with increasing redshifts, since quenching feedback
($\zeta_{\rm quench}$) is less effective at higher redshifts. For more typical
galaxies, the stellar efficiency decreases with increasing redshifts,
owing primarily to stronger ejective feedback ($\eta$) at a given
halo mass.  This is a form of downsizing, where the peak efficiency
of star formation from baryons within halos drops with time.

In the middle panel of Figure \ref{fig:prediction}, we show the
evolution of the mass--metallicity relation over the range $7\leqslant
\log(M_*/{\rm M}_\odot)\leqslant 11.4$.  Although, all the galaxies
increase in both mass and metallicity as they evolve, the shape and
slope of their relation is not constant with time and stellar mass
\citep{2008MNRAS.385.2181F}.  The evolution can actually be
well-described by a fixed faint-end slope, with a turnover to a
flat MZR at a mass that evolves downwards with time.  Such a behavior
is also inferred from observations by \citet{2014ApJ...791..130Z},
who found a turnover mass of $10^9 {\rm M}_\odot$ at $z=0$ increasing to
$10^{9.7}{\rm M}_\odot$ at $z\sim 1$; these values are in broad agreement
with our model predictions.  In our model, such a trend continues
to higher redshifts, such that at $z\gg 2$ the galaxies within the
observationally-accessible stellar mass range mostly lie on a
power-law MZR, since the turnover mass has now exceeded $10^{11}{\rm M}_\odot$.
Measuring the evolution of the MZR at a fixed $M_*$ would thus give
results that are highly mass-dependent; from $z=2\rightarrow 0$,
at $M_*=10^9 {\rm M}_\odot$ the metallicity increases by $\times 3$, where
at $M_*=10^{10.5} {\rm M}_\odot$ it increases by only $\sim 0.1$ dex.  As
metallicity measures are pushed to lower masses and wider areas at
high redshifts by surveys such as MOSDEF, this will present an
interesting testable prediction.

The predicted evolution of the MS out to $z=6$ is shown in the right
panel of Figure \ref{fig:prediction}.  Our best-fit model predicts
that, at a given $M_*$, star formation rates are increasingly higher
at earlier times \citep{2010ApJ...718.1001B}.  We also see, at late
epochs, that the $M_*$-SFR relation shows a shallower slope at high
stellar masses, which is also seen in observations
\citep{2014ApJ...795..104W}. The evolution of the SFR at a given
$M_*$ slows but does not cease to the highest redshifts, in agreement
with observations by \cite{2013ApJ...763..129S}.  We note that our
predicted relations are for the mean SFR at any given $M_*$, but
for large galaxies this could manifest as a small number of galaxies
with high SFR while most galaxies are non-star forming; our current
model does not account for such ``duty cycle" effects.

\begin{figure}
 \centerline{
   \includegraphics[height=0.45\textwidth, angle=270]{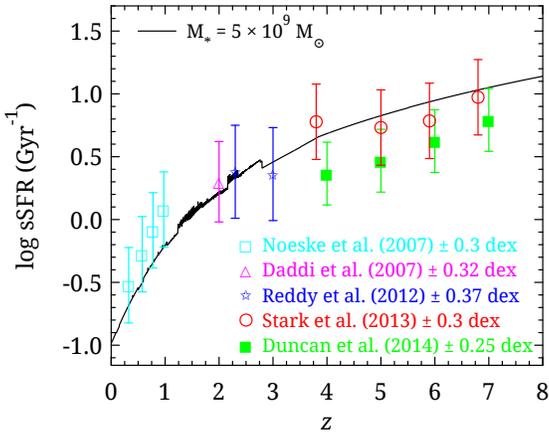}
  }
\caption{Specific SFR at $M_*=5\times 10^{9}{\rm M}_\odot$ as a function of redshift
, compared with observations listed and described further in the text.}
\label{fig:sSFR}
\end{figure}

\begin{figure}
 \centerline{
   \includegraphics[height=0.45\textwidth, angle=270]{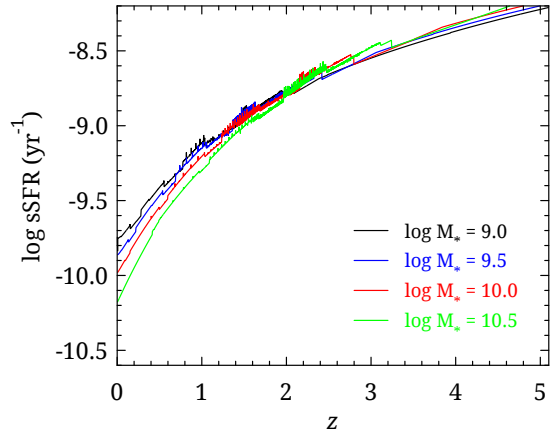}
  }
\caption{Specific SFR as a function of redshift for galaxies with different stellar masses 
of $M_*=10^{9}{\rm M}_{\odot}$ (black), $10^{9.5}{\rm M}_{\odot}$ (blue),
$10^{10}{\rm M}_{\odot}$ (red) and $10^{10.5}{\rm M}_{\odot}$ (green). Our equilibrium
model predicts somewhat different sSFR for different stellar masses at a fixed redshift.}
\label{fig:sSFRMstar}
\end{figure}

Figure~\ref{fig:sSFR} shows the specific SFR at a fixed stellar
mass of $5\times 10^9{\rm M}_\odot$ from $z=0-7$.  Observational data is
shown from \cite{2007ApJ...660L..43N,2007ApJ...670..156D,
2012ApJ...754...25R,2013ApJ...763..129S,2014MNRAS.444.2960D},
with the corresponding quoted uncertainties in the figure.
Our best fit-model underpredicts the sSFR slightly
at $z\sim 1$, but is in good agreement at $z\sim 2$ unlike in many
previous models.  At higher redshifts the observations become more
subject to systematic uncertainties particularly related to the
contribution of nebular emission lines to the flux beyond the
4000\AA\ break~\citep{2013ApJ...763..129S}.  Nonetheless, the general trend
of a slower increase in sSFR with $z$ at higher redshifts is in
broad agreement with observations, and is in good agreement
with the data from \citet{2013ApJ...763..129S} though slightly above
observations by \citet{2014MNRAS.444.2960D}.
In Figure~\ref{fig:sSFRMstar}, 
we have shown the sSFR for galaxies with different stellar masses
($M_*=10^{9}$, $10^{9.5}$, $10^{10}$ and $10^{10.5}{\rm M}_\odot$)
as a function of redshift. Unlike the widely-reported difficulty
for SAMs and hydro simulations~\citep[e.g.][]{2015MNRAS.447.3548S}
to reproduce different sSFR as a function of stellar masses, the equilibrium
model predicts somewhat mass-dependent sSFR at fixed redshift. This behavior
is in well agreement with that obtained from several empirical
models~\citep[e.g.][]{2013MNRAS.428.3121M}, at least for lower redshift range.

Overall, the predictions of the equilibrium model are in good
agreement with observations from today back to the earliest observable
epochs.  As observations at higher redshifts improve, these will
provide a more stringent test (or additional constraints) for this
model.  An additional caveat, as discussed in \cite{dav12}, is that
there is a period at very early epochs where the equilibrium
assumptions cannot be satisfied owing to inflow rates that exceed
the galaxies' ability to process that gas into stars and/or outflow.
Observationally, this ``gas accumulation epoch" is inferred to end
roughly around $z\sim 4$ \citep{2011MNRAS.412.1123P}.  We expect
the equilibrium model to begin to break down as observations probe
the gas accumulation epoch, but so far the data are insufficient
to assess this.

\subsection{Evolution of stars, gas, and metals}

A direct prediction of the equilibrium model is the growth of stars
and metals in galaxies of various masses over cosmic time.  Given
that the MCMC-constrained model correctly predicts the observed
evolution of key scaling relations, we can now explore the implications
for this.

\begin{figure}
   \includegraphics[height=0.45\textwidth, angle=270]{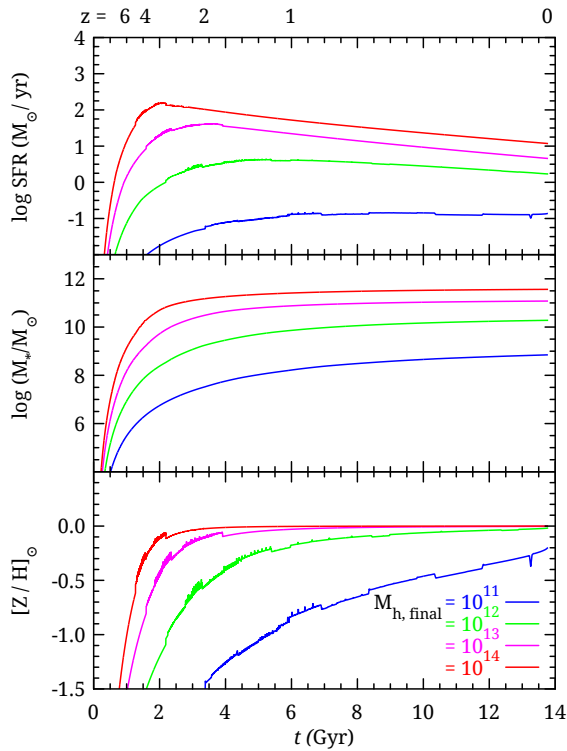}
\caption{Evolution of stars and metals of different galaxies
that have final $z=0$ halo masses of $10^{11}{\rm M}_{\odot}$ (blue), $10^{12}{\rm M}_{\odot}$ (green),
$10^{13}{\rm M}_{\odot}$ (magenta), $10^{14}{\rm M}_{\odot}$ (red). Panels from top to
bottom display the evolution of star formation rate, total stellar mass and
metallicities. The small fluctuations in the evolution of stars and metals indicate
the effect of wind recycling implemented in this model.}
\label{fig:evolution}
\end{figure}

In Figure \ref{fig:evolution}, top panel, we show the SFR evolution calculated
from our best-fit model for four galaxies that have final ($z=0$)
halo masses from $10^{11}-10^{14}{\rm M}_\odot$.  More massive galaxies
are seen to have earlier and relatively higher peak in their evolution
of SFR, consistent with the idea of downsizing \citep{2005ApJ...621..673T}.
A typical massive galaxy of $10^{14}{\rm M}_{\odot}$ halo is seen here
to have a peak in SFR of $\sim 150 - 200$ ${\rm M}_{\odot}{\rm yr^{-1}}$
at $z \sim 3-4$.  In contrast, for a Milky Way-sized halo
($10^{12}{\rm M}_{\odot}$, green lines), the SFR peaks at $\sim 3-6$
${\rm M}_{\odot}{\rm yr^{-1}}$ around $z \sim 1-2$, while at $z=0$ it has
SFR of $\sim 2$ ${\rm M}_{\odot}{\rm yr^{-1}}$, total stellar mass of
$\sim 2\times 10^{10}{\rm M}_{\odot}$ and $Z={\rm Z}_{\odot}$.  This stellar
mass is lower than observed~\citep{2015ApJ...806...96L}, which reflects the fact that our model
does not quite produce the peaked distribution in the $M_*-M_h$
relation right around $10^{12}{\rm M}_{\odot}$.  Also, if the Milky Way
halo mass is more like $2\times 10^{12}{\rm M}_{\odot}$ 
\citep{2010MNRAS.406..896B,2011MNRAS.414.2446M} as has been argued,
then the stellar mass would be commensurately higher and in better
agreement with observations.  Dwarf galaxies have increasing star
formation rates with time, but still begin forming stars at $z\gg 2$.

In the bottom panel of Figure \ref{fig:evolution}, we show
the evolution of the metallicity.  This is very rapid at the earlier
stages and then slows down
\citep{2006MNRAS.370..273D,2011ApJ...743..169F,dav12}. Once a galaxy
becomes quenched, its metallicity saturates, which in our model is
constrained to happen at around the yield value that we assume is
at solar metallicity.  Recall that this predicted metallicity
corresponds to the gas-phase oxygen abundance, which does not account
for additional enrichment of e.g. iron from Type Ia supernovae.

Note that there are small fluctuations in the evolution of stars
and metals, owing to the effect of wind recycling which takes some
of the ejected material at an early time and puts it back into the
galaxy after a time $t_{\rm rec}$.  We can also see that massive
galaxies have comparatively shorter recycling times, which is
qualitatively consistent with the simulation predictions
\citep{2008MNRAS.387..577O}.

There are many potential uses for these types of equilibrium model
evolutionary predictions.  For instance, a common technique to
obtain galaxy physical properties from broad-band data is via
spectral energy distribution (SED) fitting, which critically relies
on an assumed input star formation history (SFH).  A single galaxy's
equilibrium model run takes just seconds, so can quickly provide
realistic SFHs constrained on-the-fly while performing SED fits.
This provides greater realism compared to typical assumptions of a
constant or exponentially declining/increasing SFHs.  Another use
would be to connect progenitors of galaxies across cosmic time, by
examining the typical stellar mass growth of a galaxy observed at
a given epoch, as an alternative to using samples matched in number
densities~\citep{2013ApJ...766...33L,2013ApJ...777L..10B}.

\section{Summary}\label{sec:summary}

We present an analytic model for the growth of the stellar, metal,
and gaseous components of galaxies, based on the idea that galaxies
grow in equilibrium between inflows, outflows, and star formation,
under the assumption of a slowly-evolving gas reservoir.  This model
contains three baryon cycling parameters that represent preventive
feedback, ejective feedback, and wind recycling.  We parameterize
these as a function of redshift and halo mass, resulting in eight
free parameters that we constrain using an MCMC algorithm against
three observed scaling relations at $z=0,1,2$: (i) the stellar
mass--halo mass relation (essentially equivalent to the stellar
mass function); (ii) the mass--metallicity relation; and (iii) the
star formation rate--stellar mass relation, The resulting best-fit
model fits these observations with an overall reduced $\chi^2\approx
1.6$.  We demonstrate via Bayesian evidence that all eight parameters
provide a statistically improved fit over removing any single
parameter.  We further show that fitting any two of the three above
scaling relations results in an acceptable prediction ($\chi^2_\nu\la
2$) for the third relation.  This indicates that the equilibrium
model, with just a small number of physically meaningful parameters,
provides a good description of the mean observed evolution of the
galaxy population out to $z=2$.

Given the data we have considered, the best-fit equilibrium model 
parameters are:
\begin{eqnarray}
\eta &=& \left(\frac{M_h}{10^{10.98+0.62\sqrt{z}}}\right)^{-1.16},\\
t_{\rm rec} &=& 0.52\times10^9{\rm yr}\times(1+z)^{-0.32} \left(\frac{M_h}{10^{12}}\right)^{-0.45},\\
\zeta_{\rm quench} &=& {\rm MIN}\left[1,\left(\frac{M_h}{M_q}\right)^{-0.49}\right],\nonumber\\ 
&& \mbox{where}~~~ \frac{M_q}{10^{12} {\rm M}_\odot} = (0.96 + 0.48 z). 
\end{eqnarray}
Of course, these values may change based on the set of observations one
selects to fit, or the exact parameterisations used to describe these
quantities.  Nonetheless, the above parameters result in a good fit
to observed scaling relations out to $z\sim 2$, as well as 
predictions to higher redshifts that are broadly consistent with
observations.

There are numerous broader implications for the success of this
model.  Perhaps most significant is the simple fact that this model
framework based on simple ISM mass balance and rapid and stable gas
processing works at all.  Considering that most analytically-based
galaxy formation models have, for many decades now, all begun with
the scenario of star-forming disks forming via cooling flows from
hot gas in merging dark matter halos, it is already interesting
that one can construct a successful galaxy formation model that does
not explicitly refer to halos, cooling, merging, or a disk star
formation law.  This is not to say that such processes don't occur
or aren't important, but merely that such processes are not the
primary driver for the overall mean evolution of the galaxy population.
The equilibrium model instead forwards the idea that it is continual
smooth accretion, modulated by continual outflows that sometimes
recycle, is the dominant driver.  In a sense this model quantifies
and formalises the so-called baryon cycle, a term that has become
a fashionable buzzword but had yet to be rigorously defined.

A crucial advantage of this model is its simplicity.  With a
manageable number of physically well-defined parameters, it becomes
more straightforward to obtain physical intuition about how baryon
cycling properties impact observable galaxy properties, and vice
versa.  The small number of parameters also enables a more robust
statistical analysis, such as fitting all model parameters
simultaneously and being able to examine the Bayesian evidence to
formally characterise the necessity of each parameter.  The minimal
parameter set is in these ways an advantage over SAMs that have far
more parameters.  However, it is worth noting that SAMs concurrently
make predictions for a far larger range of galaxy properties than
the model presented here.  Hence the equilibrium model should be
viewed as a complement to SAMs, or alternatively a new and more
simpler framework on which to incorporate all the additional processes
that must be included to predict more observed aspects of the galaxy
population across cosmic time.

The model presented here only reflects the mean evolution of the
galaxy population, therefore it is a zeroeth order model for galaxy
evolution.  Fluctuations around these mean trends are likely to be
driven by the lumpy nature of dark matter that cause fluctuations
in the inflow rate, including mergers.  Inflow fluctuations can
give rise to trends such as the SFR dependence of the mass--metallicity
relation~\citep[e.g.][]{dav11b} and stellar mass--gas mass
relation~\citep[e.g.][]{2014arXiv1408.2531R}.  Moreover, the lumpiness
of inflow can affect the morphology of galaxies, by generating
dynamical instabilities within disks that grow bulges either via
mergers~\citep{1996ApJ...464..641M} or more
secularly~\citep[e.g.][]{2014MNRAS.442.1545C}.  These processes
that depend on the lumpiness of the accretion can thus be considered
to be ``first-order" galaxy evolutionary processes, which do not
drive the primary zeroeth order trends, but nonetheless are crucial
to understanding the galaxy population fully.  Indeed, the fact
that such a simple model so accurately encapsulates zeroeth order
processes suggests that this portion of the problem is now relatively
well understood, and hence at the present time the most interesting
studies of galaxy evolution may be quantifying these first-order
aspects.  A natural extension to the equilibrium model would thus
incorporate variations in the inflow rate owing to inflow fluctuations,
which can be predicted from halo assembly, but will also need to
characterise the timescale to return to equilibrium once
perturbed~\citep[e.g. the dilution time discussed
in][]{2008MNRAS.385.2181F}.  We leave this for future work.

The equilibrium model presented here explicitly does not consider
galaxy mergers.  This is in contrast to SAMs, which are based on
merger trees.  Recent models and observations have suggested that
mergers are sub-dominant for overall galaxy growth
\citep{2005MNRAS.363....2K,2007ApJ...660L..43N,2011ApJ...739L..40R}, and
our equilibrium model supports this view.  Mergers do likely drive
certain rare classes of galaxies such as local
starbursts~\citep{1996ARA&A..34..749S}, so our current model cannot
account for these.  Galaxy morphology is likewise not explicitly
considered or predicted in our model, and is likely related to
mergers.  The implication is that these aspects, while interesting,
are of secondary importance for the mean growth of stars, gas, and
metals in galaxies, though as first-order effects they will provide
interesting scatter around the mean scaling relations considered
here \citep{2010MNRAS.408.2115M}.

Environment also has important effects on galaxies.  While it appears
that central galaxy properties are primarily dependent on stellar
mass and are (to a few percent level) independent of environment,
the properties of satellites can be strongly affected by environment,
and even central galaxies that live near larger halos can be
impacted~\citep{2015MNRAS.447..374G}.  Modeling satellites will
thus require extensions of this model that properly capture additional
preventive feedback (and perhaps ejective feedback and recycling
as well) that are environment-dependent.  So far semi-analytic
models have had some difficulty reproducing the detailed properties
of satellites~\citep[e.g.][]{2010MNRAS.406.2249W,2014arXiv1404.6524H},
so this aspect of galaxy evolution remains far from solved.

Finally, we have not considered gas content in this paper, although
it can be straightforwardly incorporated into an equilibrium
model~\citep{dav12}.  This would involve introducing another
parameter, namely the gas depletion time, but this is in principle
not difficult.  The depletion time is most directly related to the
molecular gas reservoir, but as this is consumed, atomic gas can
replenish this.  Moreover, the gas reservoir within galaxies is
also intimately connected with circum-galactic and intergalactic
gas via inflows and outflows.  Given these complexities, we have
deferred considering gas reservoirs in this model to future work
that will discuss all these aspects more cohesively.  Nonetheless,
with improving observations of gas in various phases across cosmic
time, it is clear that such information will provide valuable
constraints on the baryon cycling framework of the equilibrium
model.

In summary, the equilibrium model recasts analytic galaxy formation
within a modern framework based on the simulation-driven paradigm
of baryon cycling.  Its simplicity and success at matching key
galaxy observables with a relatively small number of parameters
suggests that it captures the essence of what drives the overall
growth of galaxies.
Although, the base model has the flexibility to introduce many more
free parameters which may further improve the match between the observations
and model predictions a bit, the strength of this model is how well it does
with so few parameters and also a reduced $\chi^2$ of $1.6$ represents
a fairly acceptable fit to the data, given all the various systematic
uncertainties. We do plan to add more parameters in the
future, however, they will be designed to match additional data, such
as gas fractions and the scatter around the relations.
The MCMC framework presented here represents
a base for building models that incorporate a more diverse and
sophisticated set of physical processes including inflow fluctuations
and mergers, thereby further refining the constraints on the key
physical parameters of galaxy growth.  Future work will continue
to explore and expand upon this basic framework to better understand
the origin of an increasingly larger set of galaxy properties.

\section*{Acknowledgements}
The authors acknowledge helpful discussions with P. Behroozi, S. Lilly,
Y. Lu, N. Katz, B. Oppenheimer, and J. Zwart.  SM and RD acknowledge support
from the South African Research Chairs Initiative and the South
African National Research Foundation.  This work was supported by
the National Science Foundation under grant number AST-0847667, and
NASA grant NNX12AH86G.

\bibliography{mitra}

\begin{thebibliography}{}

\bibitem[\protect\citeauthoryear{An, Brooks \& Gelman}{An
  et~al.}{1998}]{An98stephenbrooks}
An L.,  Brooks S.,    Gelman A.,  1998, Journal of Computational and Graphical
  Statistics, 7, 434

\bibitem[\protect\citeauthoryear{{Andrews} \& {Martini}}{{Andrews} \&
  {Martini}}{2013}]{2013ApJ...765..140A}
{Andrews} B.~H.,  {Martini} P.,  2013, \apj, 765, 140

\bibitem[\protect\citeauthoryear{{Asplund}, {Grevesse}, {Sauval} \&
  {Scott}}{{Asplund} et~al.}{2009}]{2009ARA&A..47..481A}
{Asplund} M.,  {Grevesse} N.,  {Sauval} A.~J.,    {Scott} P.,  2009, Ann. Rev.
  Astron. \& Astrophys, 47, 481

\bibitem[\protect\citeauthoryear{{Behroozi}, {Marchesini}, {Wechsler},
  {Muzzin}, {Papovich} \& {Stefanon}}{{Behroozi}
  et~al.}{2013}]{2013ApJ...777L..10B}
{Behroozi} P.~S.,  {Marchesini} D.,  {Wechsler} R.~H.,  {Muzzin} A.,
  {Papovich} C.,    {Stefanon} M.,  2013, \apjl, 777, L10

\bibitem[\protect\citeauthoryear{{Behroozi}, {Wechsler} \& {Conroy}}{{Behroozi}
  et~al.}{2013}]{2013ApJ...770...57B}
{Behroozi} P.~S.,  {Wechsler} R.~H.,    {Conroy} C.,  2013, \apj, 770, 57

\bibitem[\protect\citeauthoryear{{Benson}}{{Benson}}{2014}]{2014MNRAS.444.2599%
B}
{Benson} A.~J.,  2014, \mnras, 444, 2599

\bibitem[\protect\citeauthoryear{{Benson}, {Pearce}, {Frenk}, {Baugh} \&
  {Jenkins}}{{Benson} et~al.}{2001}]{2001MNRAS.320..261B}
{Benson} A.~J.,  {Pearce} F.~R.,  {Frenk} C.~S.,  {Baugh} C.~M.,    {Jenkins}
  A.,  2001, \mnras, 320, 261

\bibitem[\protect\citeauthoryear{{Binney}}{{Binney}}{1977}]{1977ApJ...215..483%
B}
{Binney} J.,  1977, \apj, 215, 483

\bibitem[\protect\citeauthoryear{{Birrer}, {Lilly}, {Amara}, {Paranjape} \&
  {Refregier}}{{Birrer} et~al.}{2014}]{2014ApJ...793...12B}
{Birrer} S.,  {Lilly} S.,  {Amara} A.,  {Paranjape} A.,    {Refregier} A.,
  2014, \apj, 793, 12

\bibitem[\protect\citeauthoryear{{Blumenthal}, {Faber}, {Primack} \&
  {Rees}}{{Blumenthal} et~al.}{1984}]{1984Natur.311..517B}
{Blumenthal} G.~R.,  {Faber} S.~M.,  {Primack} J.~R.,    {Rees} M.~J.,  1984,
  \nat, 311, 517

\bibitem[\protect\citeauthoryear{{Bouch{\'e}} et~al.,}{{Bouch{\'e}}
  et~al.}{2010}]{2010ApJ...718.1001B}
{Bouch{\'e}} N.,  et~al., 2010, \apj, 718, 1001

\bibitem[\protect\citeauthoryear{{Bower}, {Benson} \& {Crain}}{{Bower}
  et~al.}{2012}]{2012MNRAS.422.2816B}
{Bower} R.~G.,  {Benson} A.~J.,    {Crain} R.~A.,  2012, \mnras, 422, 2816

\bibitem[\protect\citeauthoryear{{Bower}, {Benson}, {Malbon}, {Helly}, {Frenk},
  {Baugh}, {Cole} \& {Lacey}}{{Bower} et~al.}{2006}]{2006MNRAS.370..645B}
{Bower} R.~G.,  {Benson} A.~J.,  {Malbon} R.,  {Helly} J.~C.,  {Frenk} C.~S.,
  {Baugh} C.~M.,  {Cole} S.,    {Lacey} C.~G.,  2006, \mnras, 370, 645

\bibitem[\protect\citeauthoryear{{Boylan-Kolchin}, {Springel}, {White} \&
  {Jenkins}}{{Boylan-Kolchin} et~al.}{2010}]{2010MNRAS.406..896B}
{Boylan-Kolchin} M.,  {Springel} V.,  {White} S.~D.~M.,    {Jenkins} A.,  2010,
  \mnras, 406, 896

\bibitem[\protect\citeauthoryear{{Brook}, {Stinson}, {Gibson}, {Ro{\v s}kar},
  {Wadsley} \& {Quinn}}{{Brook} et~al.}{2012}]{2012MNRAS.419..771B}
{Brook} C.~B.,  {Stinson} G.,  {Gibson} B.~K.,  {Ro{\v s}kar} R.,  {Wadsley}
  J.,    {Quinn} T.,  2012, \mnras, 419, 771

\bibitem[\protect\citeauthoryear{{Brooks}, {Governato}, {Booth}, {Willman},
  {Gardner}, {Wadsley}, {Stinson} \& {Quinn}}{{Brooks}
  et~al.}{2007}]{2007ApJ...655L..17B}
{Brooks} A.~M.,  {Governato} F.,  {Booth} C.~M.,  {Willman} B.,  {Gardner}
  J.~P.,  {Wadsley} J.,  {Stinson} G.,    {Quinn} T.,  2007, \apjl, 655, L17

\bibitem[\protect\citeauthoryear{{Ceverino}, {Klypin}, {Klimek},
  {Trujillo-Gomez}, {Churchill}, {Primack} \& {Dekel}}{{Ceverino}
  et~al.}{2014}]{2014MNRAS.442.1545C}
{Ceverino} D.,  {Klypin} A.,  {Klimek} E.~S.,  {Trujillo-Gomez} S.,
  {Churchill} C.~W.,  {Primack} J.,    {Dekel} A.,  2014, \mnras, 442, 1545

\bibitem[\protect\citeauthoryear{{Chabrier}}{{Chabrier}}{2003}]{2003PASP..115.%
.763C}
{Chabrier} G.,  2003, \pasp, 115, 763

\bibitem[\protect\citeauthoryear{{Crain} et~al.,}{{Crain}
  et~al.}{2009}]{2009MNRAS.399.1773C}
{Crain} R.~A.,  et~al., 2009, \mnras, 399, 1773

\bibitem[\protect\citeauthoryear{{Creasey}, {Theuns} \& {Bower}}{{Creasey}
  et~al.}{2015}]{2015MNRAS.446.2125C}
{Creasey} P.,  {Theuns} T.,    {Bower} R.~G.,  2015, \mnras, 446, 2125

\bibitem[\protect\citeauthoryear{{Croton} et~al.,}{{Croton}
  et~al.}{2006}]{2006MNRAS.365...11C}
{Croton} D.~J.,  et~al., 2006, \mnras, 365, 11

\bibitem[\protect\citeauthoryear{{Daddi} et~al.,}{{Daddi}
  et~al.}{2007}]{2007ApJ...670..156D}
{Daddi} E.,  et~al., 2007, \apj, 670, 156

\bibitem[\protect\citeauthoryear{{Dalcanton}, {Spergel} \&
  {Summers}}{{Dalcanton} et~al.}{1997}]{1997ApJ...482..659D}
{Dalcanton} J.~J.,  {Spergel} D.~N.,    {Summers} F.~J.,  1997, \apj, 482, 659

\bibitem[\protect\citeauthoryear{{Dav{\'e}}}{{Dav{\'e}}}{2008}]{2008MNRAS.385.%
.147D}
{Dav{\'e}} R.,  2008, \mnras, 385, 147

\bibitem[\protect\citeauthoryear{{Dav{\'e}}, {Finlator} \&
  {Oppenheimer}}{{Dav{\'e}} et~al.}{2006}]{2006MNRAS.370..273D}
{Dav{\'e}} R.,  {Finlator} K.,    {Oppenheimer} B.~D.,  2006, \mnras, 370, 273

\bibitem[\protect\citeauthoryear{{Dav{\'e}}, {Finlator} \&
  {Oppenheimer}}{{Dav{\'e}} et~al.}{2011}]{dav11b}
{Dav{\'e}} R.,  {Finlator} K.,    {Oppenheimer} B.~D.,  2011, \mnras, 416, 1354

\bibitem[\protect\citeauthoryear{{Dav{\'e}}, {Finlator} \&
  {Oppenheimer}}{{Dav{\'e}} et~al.}{2012}]{dav12}
{Dav{\'e}} R.,  {Finlator} K.,    {Oppenheimer} B.~D.,  2012, \mnras, 421, 98

\bibitem[\protect\citeauthoryear{{Dav{\'e}}, {Katz}, {Oppenheimer}, {Kollmeier}
  \& {Weinberg}}{{Dav{\'e}} et~al.}{2013}]{dav13}
{Dav{\'e}} R.,  {Katz} N.,  {Oppenheimer} B.~D.,  {Kollmeier} J.~A.,
  {Weinberg} D.~H.,  2013, \mnras, 434, 2645

\bibitem[\protect\citeauthoryear{{Dav{\'e}}, {Oppenheimer} \&
  {Finlator}}{{Dav{\'e}} et~al.}{2011}]{dav11a}
{Dav{\'e}} R.,  {Oppenheimer} B.~D.,    {Finlator} K.,  2011, \mnras, 415, 11

\bibitem[\protect\citeauthoryear{{Dekel} et~al.,}{{Dekel}
  et~al.}{2009}]{2009Natur.457..451D}
{Dekel} A.,  et~al., 2009, \nat, 457, 451

\bibitem[\protect\citeauthoryear{{Dekel} \& {Mandelker}}{{Dekel} \&
  {Mandelker}}{2014}]{2014MNRAS.444.2071D}
{Dekel} A.,  {Mandelker} N.,  2014, \mnras, 444, 2071

\bibitem[\protect\citeauthoryear{{Duncan} et~al.,}{{Duncan}
  et~al.}{2014}]{2014MNRAS.444.2960D}
{Duncan} K.,  et~al., 2014, \mnras, 444, 2960

\bibitem[\protect\citeauthoryear{{Fall} \& {Efstathiou}}{{Fall} \&
  {Efstathiou}}{1980}]{1980MNRAS.193..189F}
{Fall} S.~M.,  {Efstathiou} G.,  1980, \mnras, 193, 189

\bibitem[\protect\citeauthoryear{{Faucher-Gigu{\`e}re}, {Kere{\v s}} \&
  {Ma}}{{Faucher-Gigu{\`e}re} et~al.}{2011}]{2011MNRAS.417.2982F}
{Faucher-Gigu{\`e}re} C.-A.,  {Kere{\v s}} D.,    {Ma} C.-P.,  2011, \mnras,
  417, 2982

\bibitem[\protect\citeauthoryear{{Faucher-Gigu{\`e}re}, {Quataert} \&
  {Hopkins}}{{Faucher-Gigu{\`e}re} et~al.}{2013}]{2013MNRAS.433.1970F}
{Faucher-Gigu{\`e}re} C.-A.,  {Quataert} E.,    {Hopkins} P.~F.,  2013, \mnras,
  433, 1970

\bibitem[\protect\citeauthoryear{{Feroz} \& {Hobson}}{{Feroz} \&
  {Hobson}}{2008}]{2008MNRAS.384..449F}
{Feroz} F.,  {Hobson} M.~P.,  2008, \mnras, 384, 449

\bibitem[\protect\citeauthoryear{{Feroz}, {Hobson} \& {Bridges}}{{Feroz}
  et~al.}{2009}]{2009MNRAS.398.1601F}
{Feroz} F.,  {Hobson} M.~P.,    {Bridges} M.,  2009, \mnras, 398, 1601

\bibitem[\protect\citeauthoryear{{Feroz}, {Hobson}, {Cameron} \&
  {Pettitt}}{{Feroz} et~al.}{2013}]{2013arXiv1306.2144F}
{Feroz} F.,  {Hobson} M.~P.,  {Cameron} E.,    {Pettitt} A.~N.,  2013,
  arXiv:1306.2144

\bibitem[\protect\citeauthoryear{{Finlator} \& {Dav{\'e}}}{{Finlator} \&
  {Dav{\'e}}}{2008}]{2008MNRAS.385.2181F}
{Finlator} K.,  {Dav{\'e}} R.,  2008, \mnras, 385, 2181

\bibitem[\protect\citeauthoryear{{Finlator}, {Dav{\'e}} \&
  {{\"O}zel}}{{Finlator} et~al.}{2011}]{2011ApJ...743..169F}
{Finlator} K.,  {Dav{\'e}} R.,    {{\"O}zel} F.,  2011, \apj, 743, 169

\bibitem[\protect\citeauthoryear{{Gabor} \& {Dav{\'e}}}{{Gabor} \&
  {Dav{\'e}}}{2012}]{2012MNRAS.427.1816G}
{Gabor} J.~M.,  {Dav{\'e}} R.,  2012, \mnras, 427, 1816

\bibitem[\protect\citeauthoryear{{Gabor} \& {Dav{\'e}}}{{Gabor} \&
  {Dav{\'e}}}{2015}]{2015MNRAS.447..374G}
{Gabor} J.~M.,  {Dav{\'e}} R.,  2015, \mnras, 447, 374

\bibitem[\protect\citeauthoryear{{Gnedin} \& {Kravtsov}}{{Gnedin} \&
  {Kravtsov}}{2010}]{2010ApJ...714..287G}
{Gnedin} N.~Y.,  {Kravtsov} A.~V.,  2010, \apj, 714, 287

\bibitem[\protect\citeauthoryear{{Hearin}, {Watson} \& {van den
  Bosch}}{{Hearin} et~al.}{2014}]{2014arXiv1404.6524H}
{Hearin} A.~P.,  {Watson} D.~F.,    {van den Bosch} F.~C.,  2014,
  arXiv:1404.6524

\bibitem[\protect\citeauthoryear{{Henriques}, {White}, {Thomas}, {Angulo},
  {Guo}, {Lemson}, {Springel} \& {Overzier}}{{Henriques}
  et~al.}{2014}]{2014arXiv1410.0365H}
{Henriques} B.,  {White} S.,  {Thomas} P.,  {Angulo} R.,  {Guo} Q.,  {Lemson}
  G.,  {Springel} V.,    {Overzier} R.,  2014, arXiv:1410.0365

\bibitem[\protect\citeauthoryear{{Henriques}, {White}, {Thomas}, {Angulo},
  {Guo}, {Lemson} \& {Springel}}{{Henriques}
  et~al.}{2013}]{2013MNRAS.431.3373H}
{Henriques} B.~M.~B.,  {White} S.~D.~M.,  {Thomas} P.~A.,  {Angulo} R.~E.,
  {Guo} Q.,  {Lemson} G.,    {Springel} V.,  2013, \mnras, 431, 3373

\bibitem[\protect\citeauthoryear{{Henry}, {Martin}, {Finlator} \&
  {Dressler}}{{Henry} et~al.}{2013}]{2013ApJ...769..148H}
{Henry} A.,  {Martin} C.~L.,  {Finlator} K.,    {Dressler} A.,  2013, \apj,
  769, 148

\bibitem[\protect\citeauthoryear{{Hopkins}, {Kere{\v s}}, {O{\~n}orbe},
  {Faucher-Gigu{\`e}re}, {Quataert}, {Murray} \& {Bullock}}{{Hopkins}
  et~al.}{2014}]{2014MNRAS.445..581H}
{Hopkins} P.~F.,  {Kere{\v s}} D.,  {O{\~n}orbe} J.,  {Faucher-Gigu{\`e}re}
  C.-A.,  {Quataert} E.,  {Murray} N.,    {Bullock} J.~S.,  2014, \mnras, 445,
  581

\bibitem[\protect\citeauthoryear{{Kauffmann}, {White} \&
  {Guiderdoni}}{{Kauffmann} et~al.}{1993}]{1993MNRAS.264..201K}
{Kauffmann} G.,  {White} S.~D.~M.,    {Guiderdoni} B.,  1993, \mnras, 264, 201

\bibitem[\protect\citeauthoryear{{Kere{\v s}}, {Katz}, {Weinberg} \&
  {Dav{\'e}}}{{Kere{\v s}} et~al.}{2005}]{2005MNRAS.363....2K}
{Kere{\v s}} D.,  {Katz} N.,  {Weinberg} D.~H.,    {Dav{\'e}} R.,  2005,
  \mnras, 363, 2

\bibitem[\protect\citeauthoryear{{Kewley} \& {Ellison}}{{Kewley} \&
  {Ellison}}{2008}]{2008ApJ...681.1183K}
{Kewley} L.~J.,  {Ellison} S.~L.,  2008, \apj, 681, 1183

\bibitem[\protect\citeauthoryear{{Krumholz} \& {Dekel}}{{Krumholz} \&
  {Dekel}}{2012}]{2012ApJ...753...16K}
{Krumholz} M.~R.,  {Dekel} A.,  2012, \apj, 753, 16

\bibitem[\protect\citeauthoryear{{Larson} et~al.,}{{Larson}
  et~al.}{2011}]{2011ApJS..192...16L}
{Larson} D.,  et~al., 2011, \apjs, 192, 16

\bibitem[\protect\citeauthoryear{{Leauthaud} et~al.,}{{Leauthaud}
  et~al.}{2012}]{2012ApJ...744..159L}
{Leauthaud} A.,  et~al., 2012, \apj, 744, 159

\bibitem[\protect\citeauthoryear{{Leitner} \& {Kravtsov}}{{Leitner} \&
  {Kravtsov}}{2011}]{2011ApJ...734...48L}
{Leitner} S.~N.,  {Kravtsov} A.~V.,  2011, \apj, 734, 48

\bibitem[\protect\citeauthoryear{{Leja}, {van Dokkum} \& {Franx}}{{Leja}
  et~al.}{2013}]{2013ApJ...766...33L}
{Leja} J.,  {van Dokkum} P.,    {Franx} M.,  2013, \apj, 766, 33

\bibitem[\protect\citeauthoryear{{Lewis} \& {Bridle}}{{Lewis} \&
  {Bridle}}{2002}]{2002PhRvD..66j3511L}
{Lewis} A.,  {Bridle} S.,  2002, \prd, 66, 103511

\bibitem[\protect\citeauthoryear{{Licquia} \& {Newman}}{{Licquia} \&
  {Newman}}{2015}]{2015ApJ...806...96L}
{Licquia} T.~C.,  {Newman} J.~A.,  2015, \apj, 806, 96

\bibitem[\protect\citeauthoryear{{Lilly}, {Carollo}, {Pipino}, {Renzini} \&
  {Peng}}{{Lilly} et~al.}{2013}]{2013ApJ...772..119L}
{Lilly} S.~J.,  {Carollo} C.~M.,  {Pipino} A.,  {Renzini} A.,    {Peng} Y.,
  2013, \apj, 772, 119

\bibitem[\protect\citeauthoryear{{Lu}, {Mo}, {Katz} \& {Weinberg}}{{Lu}
  et~al.}{2012}]{2012MNRAS.421.1779L}
{Lu} Y.,  {Mo} H.~J.,  {Katz} N.,    {Weinberg} M.~D.,  2012, \mnras, 421, 1779

\bibitem[\protect\citeauthoryear{{Mannucci}, {Cresci}, {Maiolino}, {Marconi} \&
  {Gnerucci}}{{Mannucci} et~al.}{2010}]{2010MNRAS.408.2115M}
{Mannucci} F.,  {Cresci} G.,  {Maiolino} R.,  {Marconi} A.,    {Gnerucci} A.,
  2010, \mnras, 408, 2115

\bibitem[\protect\citeauthoryear{{McMillan}}{{McMillan}}{2011}]{2011MNRAS.414.%
2446M}
{McMillan} P.~J.,  2011, \mnras, 414, 2446

\bibitem[\protect\citeauthoryear{{Mihos} \& {Hernquist}}{{Mihos} \&
  {Hernquist}}{1996}]{1996ApJ...464..641M}
{Mihos} J.~C.,  {Hernquist} L.,  1996, \apj, 464, 641

\bibitem[\protect\citeauthoryear{{Mo}, {Mao} \& {White}}{{Mo}
  et~al.}{1998}]{1998MNRAS.295..319M}
{Mo} H.~J.,  {Mao} S.,    {White} S.~D.~M.,  1998, \mnras, 295, 319

\bibitem[\protect\citeauthoryear{{Moster}, {Naab} \& {White}}{{Moster}
  et~al.}{2013}]{2013MNRAS.428.3121M}
{Moster} B.~P.,  {Naab} T.,    {White} S.~D.~M.,  2013, \mnras, 428, 3121

\bibitem[\protect\citeauthoryear{{Munshi}, {Governato}, {Brooks},
  {Christensen}, {Shen}, {Loebman}, {Moster}, {Quinn} \& {Wadsley}}{{Munshi}
  et~al.}{2013}]{2013ApJ...766...56M}
{Munshi} F.,  {Governato} F.,  {Brooks} A.~M.,  {Christensen} C.,  {Shen} S.,
  {Loebman} S.,  {Moster} B.,  {Quinn} T.,    {Wadsley} J.,  2013, \apj, 766,
  56

\bibitem[\protect\citeauthoryear{{Murray}, {Quataert} \& {Thompson}}{{Murray}
  et~al.}{2010}]{2010ApJ...709..191M}
{Murray} N.,  {Quataert} E.,    {Thompson} T.~A.,  2010, \apj, 709, 191

\bibitem[\protect\citeauthoryear{{Narayanan} \& {Dav{\'e}}}{{Narayanan} \&
  {Dav{\'e}}}{2013}]{2013MNRAS.436.2892N}
{Narayanan} D.,  {Dav{\'e}} R.,  2013, \mnras, 436, 2892

\bibitem[\protect\citeauthoryear{{Noeske} et~al.,}{{Noeske}
  et~al.}{2007}]{2007ApJ...660L..43N}
{Noeske} K.~G.,  et~al., 2007, \apjl., 660, L43

\bibitem[\protect\citeauthoryear{{Okamoto}, {Gao} \& {Theuns}}{{Okamoto}
  et~al.}{2008}]{2008MNRAS.390..920O}
{Okamoto} T.,  {Gao} L.,    {Theuns} T.,  2008, \mnras, 390, 920

\bibitem[\protect\citeauthoryear{{Oppenheimer} \& {Dav{\'e}}}{{Oppenheimer} \&
  {Dav{\'e}}}{2008}]{2008MNRAS.387..577O}
{Oppenheimer} B.~D.,  {Dav{\'e}} R.,  2008, \mnras, 387, 577

\bibitem[\protect\citeauthoryear{{Oppenheimer}, {Dav{\'e}}, {Kere{\v s}},
  {Fardal}, {Katz}, {Kollmeier} \& {Weinberg}}{{Oppenheimer}
  et~al.}{2010}]{2010MNRAS.406.2325O}
{Oppenheimer} B.~D.,  {Dav{\'e}} R.,  {Kere{\v s}} D.,  {Fardal} M.,  {Katz}
  N.,  {Kollmeier} J.~A.,    {Weinberg} D.~H.,  2010, \mnras, 406, 2325

\bibitem[\protect\citeauthoryear{{Papovich}, {Finkelstein}, {Ferguson}, {Lotz}
  \& {Giavalisco}}{{Papovich} et~al.}{2011}]{2011MNRAS.412.1123P}
{Papovich} C.,  {Finkelstein} S.~L.,  {Ferguson} H.~C.,  {Lotz} J.~M.,
  {Giavalisco} M.,  2011, \mnras, 412, 1123

\bibitem[\protect\citeauthoryear{{Peng} \& {Maiolino}}{{Peng} \&
  {Maiolino}}{2014}]{2014MNRAS.443.3643P}
{Peng} Y.-j.,  {Maiolino} R.,  2014, \mnras, 443, 3643

\bibitem[\protect\citeauthoryear{{Rafieferantsoa}, {Dav{\'e}},
  {Angl{\'e}s-Alcazar}, {Katz}, {Kollmeier} \& {Oppenheimer}}{{Rafieferantsoa}
  et~al.}{2014}]{2014arXiv1408.2531R}
{Rafieferantsoa} M.,  {Dav{\'e}} R.,  {Angl{\'e}s-Alcazar} D.,  {Katz} N.,
  {Kollmeier} J.~A.,    {Oppenheimer} B.~D.,  2014, arXiv:1408.2531

\bibitem[\protect\citeauthoryear{{Reddy}, {Pettini}, {Steidel}, {Shapley},
  {Erb} \& {Law}}{{Reddy} et~al.}{2012}]{2012ApJ...754...25R}
{Reddy} N.~A.,  {Pettini} M.,  {Steidel} C.~C.,  {Shapley} A.~E.,  {Erb} D.~K.,
     {Law} D.~R.,  2012, \apj, 754, 25

\bibitem[\protect\citeauthoryear{{Rees} \& {Ostriker}}{{Rees} \&
  {Ostriker}}{1977}]{1977MNRAS.179..541R}
{Rees} M.~J.,  {Ostriker} J.~P.,  1977, \mnras, 179, 541

\bibitem[\protect\citeauthoryear{{Rodighiero} et~al.,}{{Rodighiero}
  et~al.}{2011}]{2011ApJ...739L..40R}
{Rodighiero} G.,  et~al., 2011, \apjl., 739, L40

\bibitem[\protect\citeauthoryear{{Saintonge} et~al.,}{{Saintonge}
  et~al.}{2013}]{2013ApJ...778....2S}
{Saintonge} A.,  et~al., 2013, \apj, 778, 2

\bibitem[\protect\citeauthoryear{{Salim} et~al.,}{{Salim}
  et~al.}{2007}]{2007ApJS..173..267S}
{Salim} S.,  et~al., 2007, \apjs, 173, 267

\bibitem[\protect\citeauthoryear{{Sanders} \& {Mirabel}}{{Sanders} \&
  {Mirabel}}{1996}]{1996ARA&A..34..749S}
{Sanders} D.~B.,  {Mirabel} I.~F.,  1996, \araa, 34, 749

\bibitem[\protect\citeauthoryear{{Sanders} et~al.,}{{Sanders}
  et~al.}{2015}]{2015ApJ...799..138S}
{Sanders} R.~L.,  et~al., 2015, \apj, 799, 138

\bibitem[\protect\citeauthoryear{{Schaye} et~al.,}{{Schaye}
  et~al.}{2015}]{2015MNRAS.446..521S}
{Schaye} J.,  et~al., 2015, \mnras, 446, 521

\bibitem[\protect\citeauthoryear{{Schreiber} et~al.,}{{Schreiber}
  et~al.}{2015}]{2015A&A...575A..74S}
{Schreiber} C.,  et~al., 2015, \aap, 575, A74

\bibitem[\protect\citeauthoryear{{Somerville} \& {Dav{\'e}}}{{Somerville} \&
  {Dav{\'e}}}{2015}]{Somerville:2015}
{Somerville} R.~S.,  {Dav{\'e}} R.,  2015, Submitted to ARA\&A

\bibitem[\protect\citeauthoryear{{Somerville}, {Hopkins}, {Cox}, {Robertson} \&
  {Hernquist}}{{Somerville} et~al.}{2008}]{2008MNRAS.391..481S}
{Somerville} R.~S.,  {Hopkins} P.~F.,  {Cox} T.~J.,  {Robertson} B.~E.,
  {Hernquist} L.,  2008, \mnras, 391, 481

\bibitem[\protect\citeauthoryear{{Somerville} \& {Primack}}{{Somerville} \&
  {Primack}}{1999}]{1999MNRAS.310.1087S}
{Somerville} R.~S.,  {Primack} J.~R.,  1999, \mnras, 310, 1087

\bibitem[\protect\citeauthoryear{{Sparre} et~al.,}{{Sparre}
  et~al.}{2015}]{2015MNRAS.447.3548S}
{Sparre} M.,  et~al., 2015, \mnras, 447, 3548

\bibitem[\protect\citeauthoryear{{Speagle}, {Steinhardt}, {Capak} \&
  {Silverman}}{{Speagle} et~al.}{2014}]{2014ApJS..214...15S}
{Speagle} J.~S.,  {Steinhardt} C.~L.,  {Capak} P.~L.,    {Silverman} J.~D.,
  2014, \apjs, 214, 15

\bibitem[\protect\citeauthoryear{{Stark}, {Schenker}, {Ellis}, {Robertson},
  {McLure} \& {Dunlop}}{{Stark} et~al.}{2013}]{2013ApJ...763..129S}
{Stark} D.~P.,  {Schenker} M.~A.,  {Ellis} R.,  {Robertson} B.,  {McLure} R.,
   {Dunlop} J.,  2013, \apj, 763, 129

\bibitem[\protect\citeauthoryear{{Steidel} et~al.,}{{Steidel}
  et~al.}{2014}]{2014ApJ...795..165S}
{Steidel} C.~C.,  et~al., 2014, \apj, 795, 165

\bibitem[\protect\citeauthoryear{{Tacconi} et~al.,}{{Tacconi}
  et~al.}{2013}]{2013ApJ...768...74T}
{Tacconi} L.~J.,  et~al., 2013, \apj, 768, 74

\bibitem[\protect\citeauthoryear{{Thomas}, {Maraston}, {Bender} \& {Mendes de
  Oliveira}}{{Thomas} et~al.}{2005}]{2005ApJ...621..673T}
{Thomas} D.,  {Maraston} C.,  {Bender} R.,    {Mendes de Oliveira} C.,  2005,
  \apj, 621, 673

\bibitem[\protect\citeauthoryear{{Torrey}, {Vogelsberger}, {Genel}, {Sijacki},
  {Springel} \& {Hernquist}}{{Torrey} et~al.}{2014}]{2014MNRAS.438.1985T}
{Torrey} P.,  {Vogelsberger} M.,  {Genel} S.,  {Sijacki} D.,  {Springel} V.,
  {Hernquist} L.,  2014, \mnras, 438, 1985

\bibitem[\protect\citeauthoryear{{Tremonti} et~al.,}{{Tremonti}
  et~al.}{2004}]{2004ApJ...613..898T}
{Tremonti} C.~A.,  et~al., 2004, \apj, 613, 898

\bibitem[\protect\citeauthoryear{{Vogelsberger} et~al.,}{{Vogelsberger}
  et~al.}{2014}]{2014Natur.509..177V}
{Vogelsberger} M.,  et~al., 2014, \nat, 509, 177

\bibitem[\protect\citeauthoryear{{Weinberg}}{{Weinberg}}{2009}]{2009arXiv0911.%
1777W}
{Weinberg} M.~D.,  2009, arXiv:0911.1777

\bibitem[\protect\citeauthoryear{{Weinmann}, {Kauffmann}, {von der Linden} \&
  {De Lucia}}{{Weinmann} et~al.}{2010}]{2010MNRAS.406.2249W}
{Weinmann} S.~M.,  {Kauffmann} G.,  {von der Linden} A.,    {De Lucia} G.,
  2010, \mnras, 406, 2249

\bibitem[\protect\citeauthoryear{{Weinmann}, {Pasquali}, {Oppenheimer},
  {Finlator}, {Mendel}, {Crain} \& {Macci{\`o}}}{{Weinmann}
  et~al.}{2012}]{2012MNRAS.426.2797W}
{Weinmann} S.~M.,  {Pasquali} A.,  {Oppenheimer} B.~D.,  {Finlator} K.,
  {Mendel} J.~T.,  {Crain} R.~A.,    {Macci{\`o}} A.~V.,  2012, \mnras, 426,
  2797

\bibitem[\protect\citeauthoryear{{Whitaker} et~al.,}{{Whitaker}
  et~al.}{2014}]{2014ApJ...795..104W}
{Whitaker} K.~E.,  et~al., 2014, \apj, 795, 104

\bibitem[\protect\citeauthoryear{{White} \& {Frenk}}{{White} \&
  {Frenk}}{1991}]{1991ApJ...379...52W}
{White} S.~D.~M.,  {Frenk} C.~S.,  1991, \apj, 379, 52

\bibitem[\protect\citeauthoryear{{White} \& {Rees}}{{White} \&
  {Rees}}{1978}]{1978MNRAS.183..341W}
{White} S.~D.~M.,  {Rees} M.~J.,  1978, \mnras, 183, 341

\bibitem[\protect\citeauthoryear{{Zahid}, {Dima}, {Kudritzki}, {Kewley},
  {Geller}, {Hwang}, {Silverman} \& {Kashino}}{{Zahid}
  et~al.}{2014}]{2014ApJ...791..130Z}
{Zahid} H.~J.,  {Dima} G.~I.,  {Kudritzki} R.-P.,  {Kewley} L.~J.,  {Geller}
  M.~J.,  {Hwang} H.~S.,  {Silverman} J.~D.,    {Kashino} D.,  2014, \apj, 791,
  130

\end{thebibliography}
\bibliographystyle{mn2e}

\renewcommand{\theequation}{A\arabic{equation}}
\setcounter{equation}{0}  
\section*{APPENDIX: Equilibrium relations}  

Star-forming galaxies closely follow the slowly evolving equilibrium condition
between the accretion of cold gas, consumption of gas into stars and outflows.
This can be obtained from the basic conservation equation for gas
mass \citep{2008MNRAS.385.2181F,2010ApJ...718.1001B,2014MNRAS.444.2071D}:
\begin{equation}\label{eqn:conserv}
\dot{M}_{\rm gas} = \dot{M}_{\rm in} - \dot{M}_{\rm *} - \dot{M}_{\rm out}
\end{equation}
where $\dot{M}_{\rm in}$ is the gas accretion rate onto the galaxy's
star-forming region, $\dot{M}_{\rm *}$ is star formation rate (SFR),
and $\dot{M}_{\rm out}$ is the gas outflow rate.
Now, as suggested by hydrodynamic simulations
\citep{2008MNRAS.385.2181F} and analytical studies
of galaxy's self-regulating
behavior to a steady state \citep{2010ApJ...718.1001B,2012ApJ...753...16K},
along with observations that show that
the gas content of galaxies evolves much more slowly than the star
formation rate \citep{2013ApJ...768...74T,2013ApJ...778....2S},
we make the {\it equilibrium assumption} that 
$\dot{M}_{\rm gas}\simeq0$ when compared to inflow rates, outflow rates, and
SFR \citep{dav12}.  This yields
\begin{equation}\label{eqn:equilnew}
 \dot{M}_{\rm in} = \dot{M}_{\rm out} + \dot{M}_{\rm *} = (1+\eta)\dot{M}_{\rm *},
\end{equation}
where $\eta\equiv\dot{M}_{\rm out}/\dot{M}_{\rm *}$ is the outflow
{\it mass loading factor}.
$\dot{M}_{\rm in}$ is governed by a combination of (i) the
gravitational-driven baryonic inflow of dark matter halos ($\dot{M}_{\rm
grav}$) (ii) the gas entering the halo that is prevented from
reaching the ISM ($\dot{M}_{\rm prev}$) and (iii) the gas recycling
rate ($\dot{M}_{\rm recyc}$), associated with the fraction of mass
that was ejected earlier in outflows, as follows:
\begin{equation}\label{eqn:Mdotin}
 \dot{M}_{\rm in} = \dot{M}_{\rm grav}-\dot{M}_{\rm prev}+\dot{M}_{\rm recyc}
 = \zeta\dot{M}_{\rm grav}+\dot{M}_{\rm recyc}
\end{equation}
where, $\zeta\equiv1-\dot{M}_{\rm prev}/\dot{M}_{\rm grav}$ is the
{\it preventive feedback parameter}.
Combining Equation \ref{eqn:equilnew} and \ref{eqn:Mdotin},
we get \citep{dav12}
\begin{equation}\label{eqn:SFRnew}
\dot{M}_{\rm *} = \frac{\zeta\dot{M}_{\rm grav}+\dot{M}_{\rm recyc}}{1+\eta}
\end{equation}
Here $\dot{M}_{\rm grav}$ is taken from \cite{2009Natur.457..451D} as
\begin{equation}\label{eqn:Mhdot}
\frac{\dot{M}_{\rm grav}}{M_h}
= 0.47 f_b \Bigl(\frac{M_h}{10^{12} {\rm M}_\odot}\Bigr)^{0.15} \Bigl(\frac{1+z}{3}\Bigr)^{2.25}\;{\rm Gyr}^{-1}.
\end{equation}
and $\zeta$ can be written as a combination of various contributions,
described in Equation~\ref{eqn:zeta} of the main text,
coming from different sources. For those, we take
(following Equation 1 of \citealt{2008MNRAS.390..920O})
\begin{equation}
 \zeta_{\rm photo}=\left[1+\frac{1}{3}\left(\frac{M_h}{M_\gamma}\right)^{-2}\right]^{-1.5}
\end{equation}
with $M_\gamma$ given in Figure 3 of \cite{2008MNRAS.390..920O}.
For $\zeta_{\rm winds}$, we consider a somewhat arbitrary parameterization
to incorporate its qualitative effects \citep{dav12,2013MNRAS.436.2892N}:
\begin{equation}
  \zeta_{\rm winds} = 1-e^{-\sqrt{M_h/M_w}},\mbox{~}
  M_w = 10^{10-0.25 z}
\end{equation}
and we take $\zeta_{\rm grav}$ as derived from hydrodynamic simulations 
with no outflows \citep{2011MNRAS.417.2982F},
\begin{equation}\label{eqn:zetagrav}
\zeta_{\rm grav} = 0.47\Bigl(\frac{1+z}{4}\Bigr)^{0.38} 
\Bigl(\frac{M_h}{10^{12}{\rm M}_\odot}\Bigr)^{-0.25}.
\end{equation}

To follow $\dot{M}_{\rm recyc}$ in our model, we use the recycling
time ($t_{\rm rec}$) information gained from
simulations \citep{2008MNRAS.387..577O,2013MNRAS.431.3373H} to track
recycling directly. The idea is that, whenever a mass is ejected
via outflows, we take that mass (i.e. $\eta\times{\rm SFR}\times{\rm d}t$)
in a given time-step ${\rm d}t$ and put it into a ``reservoir" as
$M_{\rm recyc}^{\rm res}$. Then after some time $t_{\rm rec}$, the
material is re-accreted into the galaxy at a rate
$\dot{M}_{\rm recyc}=M_{\rm recyc}^{\rm res}/{\rm d}t$, where ${\rm d}t$
is the time-step at time of re-accretion. Note that, here ${\rm d}t$
is not constant; we adjust it to keep $0.001<{\rm d}M_h/M_h<0.01$,
where ${\rm d}M_h = \dot{M}_h\times{\rm d}t$.

We also include mass loss from stellar
evolution $R(t)$ that goes back into gas, so that the net SFR becomes
$(1-R)\dot{M}_{\rm *}$, where $R(t)$ for a
Chabrier \citep{2003PASP..115..763C} IMF is well-approximated
by \citep{2011ApJ...734...48L,2013MNRAS.428.3121M}
\begin{equation}
 R(t) = 0.046\ln\left(\frac{t}{2.76\times10^5 {\rm~yr}}+1\right)
\end{equation}

Now, the metal content of galaxies, $Z_{\rm ISM}$ can be derived
from the yield times SFR divided by the inflow rate which is yet
to be enriched:
\begin{equation}\label{eqn:Znew}
Z_{\rm ISM} = \frac{y\dot{M}_{\rm *}}{\dot{M}_{\rm in}-\dot{M}_{\rm recyc}}
= \frac{y\dot{M}_{\rm *}}{\zeta\dot{M}_{\rm grav}}.
\end{equation}
This assumes that the metallicity of re-accreted outflows is similar
to that of the galaxy; in detail, the metallicity of a galaxy
increases slightly with time, but the outflowing gas is also likely
slightly enriched relative to the galaxy at the time of ejection.

The above Equations \ref{eqn:SFRnew} and \ref{eqn:Znew} are usually
referred to as the {\it equilibrium relations} and the parameters
$\eta$, $\zeta$ and $t_{\rm rec}$ describing these relations are
known as {\it baryon cycling parameters}.  Constraining these
parameters based on various observations is the main goal of this
work.

\end{document}